\documentclass{aastex61}

\received{October 27, 2017}
\revised{December 14, 2017}
\accepted{\today}
\submitjournal{ApJ}

\shorttitle{novae late panchromatic spectra}
\shortauthors{Mason et al.}

\begin{document}

\title{V1369 Cen high resolution panchromatic late nebular spectra in the context of a unified picture for nova ejecta}

\correspondingauthor{Elena Mason}
\email{emason@oats.inaf.it}

\author{Elena Mason}
\affil{INAF-OATS, Via G.B. Tiepolo, 11, 34143, Trieste, IT}

\author{Steven N. Shore}
\affiliation{Dipartimento di Fisica ``Enrico Fermi'', Universita' di Pisa}
\affiliation{INFN Pisa, Largo B. Pontecorvo, 56127, PI, IT}
\affiliation{Astronomical Institute, Charles University in Prague, V Hole\v{s}ovi\v{c}k\'ach 2, 180 00, Praha 8, CZ}

\author{Ivan De Gennaro Aquino}
\affiliation{Hamburger Sternwarte, Gojenbergsweg 112, 21029 Hamburg, DE}

\author{Luca Izzo}
\affiliation{Instituto de Astrofisica de Andalucia, Glorieta de la Astronomia s/n, E-18008 Granada, ES}

\author{Kim Page}
\affiliation{Department of Physics and Astronomy, University of Leicester, Leicester, LE1 7RH, UK}

\author{Greg J. Schwarz}
\affiliation{American Astronomical Society, 2000 Florida Ave., NW, Suite 400, DC 20009-1231, USA}



\begin{abstract}

Nova Cen 2013 (V1369 Cen) is the fourth bright nova observed panchromatically through high resolution UV+optical multi epoch spectroscopy. It is also the nova with the richest set of spectra (both in terms of data quality and number of epochs) thanks to its exceptional brightness. Here, we use the late nebular spectra taken between day $\sim$250 and day $\sim$837 after outburst to derive the physical, geometrical and kinematical properties of the nova. We compare the results with those determined for the other panchromatic studies in this series: T Pyx, V339 Del (nova Del 2013), and V959 Mon (nova Mon 2012). From this we conclude that in all these novae the ejecta geometry and phenomenology can be consistently explained by clumpy gas expelled during a single, brief ejection episode and in ballistic expansion, and not by a wind. For  V1369 Cen the ejecta mass ($\sim1\times$10$^{-4}$ M$\odot$) and filling factor (0.1$\leq f \leq$0.2) are consistent with those of classical novae but larger (by at least an order of magnitude) than those of T Pyx and the recurrent novae. V1369 Cen has an anomalously high relative to solar N/C ratio that is beyond the range currently predicted for a CO nova, and the Ne emission line strengths are dissimilar to those of typical ONe or CO white dwarfs.

\end{abstract}




\section{Introduction} \label{sec:intro}

For classical and recurrent novae, the geometry of the ejecta, and its connection to the explosion, the ejection mechanism, the distance and the ejecta mass, has been investigated by several authors. There are various approaches to constrain the geometry. They can be purely geometrical (e.g. Payne-Gaposchkin 1957,  Gill \& O'Brien 1999, Gill \& O'Brien 2000, Banerjee et al. 2016), or they can include  kinematics (e.g. Chochol et al. 1997; Harman \& O'Brien 2003; Eyres et al. 2005;  Ribeiro et al. 2011, Munari et al. 2011, Ribeiro et al. 2013, Lindford et al. 2015, Harvey et al. 2016). These assume geometrical components to start with, and choose them in combination to fit a model to the observations. In the most sophisticated cases, the best fit indicates components, aspect ratio, maximum velocity and also mass distribution of the observed ejecta. The dynamics are usually adopted from independent studies, and so is the ejection mechanism. 

Attempts to identify the ejection mechanism and solve the long standing debate upon the windy or shell nature of the ejecta include, for example, those of Payne-Gaposchkin \& Menzel (1938), Payne-Gaposchkin (1957), Friedjung (1966a, 1966b), Hutchings (1970), Hjellming (1990), Williams (1992), Scott et al. (1995), Williams et al. (1996), Cassatella et al. (2004a), Liimets et al. (2012). However, without adequate  theoretical modeling of the thermonuclear ignition and the ejecta lift-off to confront with observations, any proposed scenario is potentially equally valid. 
Since nova ejecta are an expanding medium at relatively high velocity, their physical conditions are constantly changing and out of equilibrium. A successful scenario therefore should explain the observations from the rise phase to the very late decline. In addition, it should be able to make sense of multi wavelength observations across the whole electromagnetic spectrum. 
With this in mind we have analyzed and {\it compared} multi-epoch panchromatic high resolution spectroscopic observations for a number of recent bright novae that in view of their brightness could be followed up to years after outburst. 

The use of high resolution spectroscopy (R$\geq$40000) was motivated by the evidence that nova ejecta display clumpy structures. Evidence for clumps has been presented, for example, by early high resolution photographic spectroscopy (e.g. Payne-Gaposchkin 1957; McLaughlin 1957, 1964), high resolution IUE observations (e.g. Cassatella et al. 2004b and references therein) and resolved images of old nova remnants (e.g. Gill \& O'Brien 2000, Harman \& O'Brien 2003, Liimets et al. 2012 and references therein, Shara et al. 2015, see also O'Brien \& Bode 2008 and references therein).
The effort to obtain panchromatic observations (i.e. contemporaneous, when not simultaneous, UV and optical spectra) was motivated by the need to account for the ionization structure of the ejecta. 

Last but not least, because of relatively sparse spectroscopy in the literature of novae in the late nebular stage, we made sure to monitor each of the sufficiently bright novae well after outburst and the super soft source (SSS) phase.
An extended monitoring became compelling after the T~Pyx 2011 outburst, which was followed from day 2 to 834, revealed stationary absorption structures in the UV resonant transitions of the late nebular spectra. Those stationary absorptions, together with the analysis of the pseudo-P-Cyg absorptions during the nova early decline, showed that the ejecta are in ballistic expansion and consistent with a biconical geometry (Shore et al. 2011, 2013a, De Gennaro Aquino et al. 2014). Hence, panchromatic Target of Opportunity (ToO) and long-term monitoring campaigns were organized to follow any bright nova through high resolution UV+optical spectroscopy. Complementary Swift observations and photometric monitoring (AAVSO, Kafka 2017) were used to program the cadence and the exposure time of the observations. The bright novae that could be followed by our programs were V339 Del (nova~Del~2013), V1369 Cen (nova~Cen~2013) and in part V959 Mon (nova Mon 2012).

Partial results have already been published for V959 Mon (Shore et al. 2013b) and V339 Del (Shore et al. 2016) showing that the biconical geometry can describe those two objects as well. In addition, early spectra of V339 Del (V959 Mon was discovered during its nebular phase) displayed absorption structures whose development was quite similar to those of T Pyx (Shore et al. 2011)

This paper presents an analysis of the late nebular spectra of V1369 Cen and compares the results with the late nebular spectra collected for the panchromatic novae listed above and not yet published. The major conclusion of this paper goes beyond the physical parameters derived for the ejecta of V1369 Cen and underlines the commonalities among the individual spectroscopic evolutions. 

Every individual object for which we have collected comparable data sets is essentially different in its physical parameters, development time scale and detailed line profiles. However, from their comparative analysis it emerges that a single assumption (the single explosion) allows us to describe each phenomenon within a single unified scenario: the visible (at any time and any wavelength) nova ejecta are not a wind but are rather consistent with an ensemble of structures in ballistic or free expansion (v$\sim$r, also called a Hubble flow) confined within a biconical geometry.

The paper consists of several parts. The first part presents the data collection and selection (Sections 2 and 3). The second part uses the data to characterize V1369 Cen quantitatively, describing its spectral evolution and deriving all the possible physical parameters of the nova (Sections 4 and 5). Section 6 analyzes transitions and line profiles to constrain the ejecta ionization structure and geometry. The line profile analysis is exemplified using V1369 Cen spectra but the results are presented together with those derived for T~Pyx, V339 Del and V959 Mon in the previous publications.
We discuss the implication of our result in Section 7. We will show in particular that our proposed scenario is not only consistent with each of the collected high resolution panchromatic epochs (from day 0 to, virtually, the undetectability), but also with recent results from other regimes of the electromagnetic spectrum. Summary and conclusions are in Section 8.

\section{Observations and data reduction} \label{sec:obs}
\subsection{Panchromatic high resolution spectroscopy}

\begin{deluxetable*}{cccccc|ccccc}[b!]\label{tab:log_cen}
\tablecaption{Log of observations for nova Cen 2013. For each instrument/instrument-setup the total integration time is given in seconds. The nova age is given in days since outburst/discovery (T$_0$ is Dec 2.692UT 2013 or MJD=56628.69). }
\tablenum{1}
\tablewidth{0pt}
\scriptsize
\tablehead{
\multicolumn{6}{c}{STIS (UV)} & \multicolumn{5}{c}{OPT instrument} \\
\colhead{UT date}   & \colhead{age}  & \colhead{E140M/1425} & \colhead{E230M/1978} & \colhead{E230M/2707} & \colhead{prop. ID} & \colhead{UT date} & \colhead{age} & \colhead{FEROS} & \colhead{UVES} & \colhead{prop. ID} \\
\colhead{} & \colhead{(d)}  &  \colhead{} &  \colhead{} & \colhead{} & \colhead{} & \colhead{} &  \colhead{(d)} &  \colhead{} & \colhead{DIC1+DIC2} & \colhead{}
}
\startdata
2014-08-12 & 252.7 &  249$\dagger$ & 249 & 249 & 13388 & 2014-08-02 & 242.4 & 900 & - & 094.A-9011 \\
2014-10-03$\ddagger$ & 304.9  & 249$\dagger$ & 249 & 249 & 13388 & -  & -& -& - & -  \\
2015-03-06 & 458.4 & 800 & 602 & 601 & 13828 & 2015-03-06 & 458.5$\ast$ & - & 15$\times$2+35$\times$3+600 & 095.D-0722 \\
           &   &     &     &     &       & 2015-04-06 & 489.6   & - & 100+180+600 &  095.D-0722 \\
2015-06-21 & 566.1 & 800 & 602 & 601 & 13828 & 2015-06-21 & 566.3  & - & 100+180+600 &  095.D-0722 \\
2016-03-19 & 837.7 & 2679 & 1440 & 1439 & 14338 &  2016-03-17 & 835.5 &  - & 100$\times$2+200+900 & 096.D-0226 \\
\enddata
\tablecomments{\\ $\dagger$ In this case the FUV setup was grism E140H centered at 1234, 1416 and 1598 \AA. Each setup had exposure 249 s. The spectral resolution of the E140H is R=114000.\\
$\ddagger$ The object was not visible from ground at the time of the HST observations. \\
$\ast$ Low signal to noise data therefore not considered for quantitative measurements in the text/paper.}
\end{deluxetable*}

\begin{deluxetable*}{cccccc|cccc}\label{tab:log_del}
\tablecaption{Log of observations for nova Del 2013. For each instrument/instrument-setup the total integration time is given in seconds. The nova age is given in days since outburst/discovery (T$_0$ is Aug 14.58UT 2013 or MJD=56518.58).}
\tablenum{2}
\tablewidth{0pt}
\scriptsize
\tablehead{
\multicolumn{6}{c}{STIS (UV)} & \multicolumn{4}{c}{OPT instrument} \\
 \colhead{UT date}   & \colhead{age}   & \colhead{E140M/1425} & \colhead{E230M/1978} & \colhead{E230M/2707} & \colhead{prop. ID} & \colhead{UT date} & \colhead{age} &  \colhead{FIES} &  \colhead{prop. ID} \\
 \colhead{} & \colhead{(d)} & \colhead{} & \colhead{} & \colhead{} & \colhead{} & \colhead{} & \colhead{(d)} & \colhead{} & \colhead{} 
 }
\startdata 
2014-10-16 & 427.4 &  587 & 587 & 587 & 13388 & 2014-10-29 & 440 &  3600 &   P49-402 \\
2015-05-11 & 635.3 &  2455 & 1328 & 1328 & 13828 & 2015-05-27 & 651 & 2000 &  P51-103 \\
2015-09-19$\dagger$ & 765.5 &  2455 & - & - & 13828 & - & - & - & - \\
2015-12-29 & 866.4 & 2455 & 1238 & 1237 & 13828 & - & - &  - & -\\
\enddata
\tablecomments{$\dagger$ The 2015-09-19 HST visit failed (only the FUV setup could be obtained) and was repeated on December 29, 2015. At both epochs the optical spectrum could not be obtained. }
\end{deluxetable*}

\begin{deluxetable*}{cccccc|ccccc}\label{tab:log_mon}
\tablecaption{Log of observations for nova Mon 2012. For each instrument/instrument-setup the total integration time is given in seconds.}
\tablenum{3}
\tablewidth{0pt}
\scriptsize
\tablehead{
\multicolumn{6}{c}{STIS (UV)} & \multicolumn{5}{c}{OPT instrument} \\
  \colhead{UT date}   & \colhead{age}  & \colhead{E140M/1425} & \colhead{E230M/1978} & \colhead{E230M/2707} & \colhead{prop. ID} & \colhead{UT date} & \colhead{age}  & \multicolumn{2}{c}{UVES} & \colhead{prop. ID} \\
 & \colhead{(d)} & \colhead{} & \colhead{} & \colhead{} & \colhead{} & \colhead{} & \colhead{(d)} & \colhead{DIC1} & \colhead{DIC2} & \colhead{}
 }
\startdata
2015-03-3/4/7 & 988$\dagger$ &  2441$\times$2+2970$\times$2 & 2411+2970 & 2970 & 13828 & 2015-02-18 & 974$\dagger$ & 1100$\times2$+2160 & 2160 &  095.D-0722 \\
\enddata
\tablecomments{
$\dagger$ For nova Mon 2012, the age is counted from the first $\gamma$-ray detection (June 19, 2012; {\it Fermi} collaboration 2014).}
\end{deluxetable*}

For each object, the panchromatic data set consists of coordinated optical and UV spectra. Early epochs had the requirement of nearly simultaneous observations, while the late nebular spectra presented herein could be up to 1 month apart for easy scheduling. 
UV spectra were secured with the Space Telescope Imaging Spectrograph (STIS) on board of the Hubble Space Telescope (HST). The adopted instrument setup was generally the same using the medium resolution echelle gratings E140M centered at 1425 \AA\ and E230M centered at 1978 and 2707 \AA\  on consecutive exposures. The resulting UV range was 1200-2900 \AA\ with spectral resolution R=45800 and R=30000 for the FUV and NUV band, respectively. The exception was early in the V1369 Cen outburst when the nova was still too bright to be safely observed with the medium echelle in the FUV. Then the high resolution echelle was used, see Table 1.

The optical spectra were obtained either at the Nordic Optical Telescope (NOT) using the FIbre-fed Echelle Spectrograph (FIES) for the case of the northern objects (e.g. V339 Del), or with the Fiber-fed Extended Range Optical Spectrograph (FEROS) at 2.2m telescope and the Ultraviolet and Visual Echelle Spectrograph (UVES) at the Very Large Telescope (VLT) for the southern objects (V1369 Cen). The covered optical range varied slightly depending on the instrument, the largest being offered by UVES that is the only one capable of covering the whole visible band from the atmosphere cutoff at 3000 \AA\ to 10000 \AA. We took advantage of this capability by combining the dichroic setups DIC1/3460+5800 and DIC2/4370+8600 \AA. FIES and FEROS have fixed spectral format and cover, respectively, the wavelength range 4000-7500 \AA\, and 4000-9000 \AA. All spectra had high resolution: R=48000 for FEROS, R=45000 or 67000 for FIES and R=40000 in UVES. 

The detailed log of observations is provided in Tables 1-3. Note that, while performed, the V959 Mon STIS observations failed so no UV spectrum exists for the last epoch of observation of V959 Mon (day 988). 

STIS data were reduced with the instrument pipeline. The order merging was, however, performed with the HST Goddard High Resolution Spectrograph (GHRS) script because of its better performance. Instrument pipeline reductions were also used for the optical spectra. We flux calibrated FEROS and FIES data using our own scripts (IDL and/or IRAF). For each object we observed a standard star which was relatively close in pointing direction. The standard star spectra (BD+28~4211 for V339 Del at FIES and LTT4816 + CD-32~9927 for V1369 Cen at FEROS) were taken immediately after the science observation. Although we also observed dedicated standard stars for the UVES runs (HR2315 for V959 Mon and LTT4816 for V1369 Cen), we finally adopted the master response function delivered by ESO for its better response below 3500 \AA\, and above 9000 \AA. We ascertained that the master response function matched the response function built with the dedicated standard star in the overlap range. 

   \begin{figure}
   \includegraphics[width=9cm]{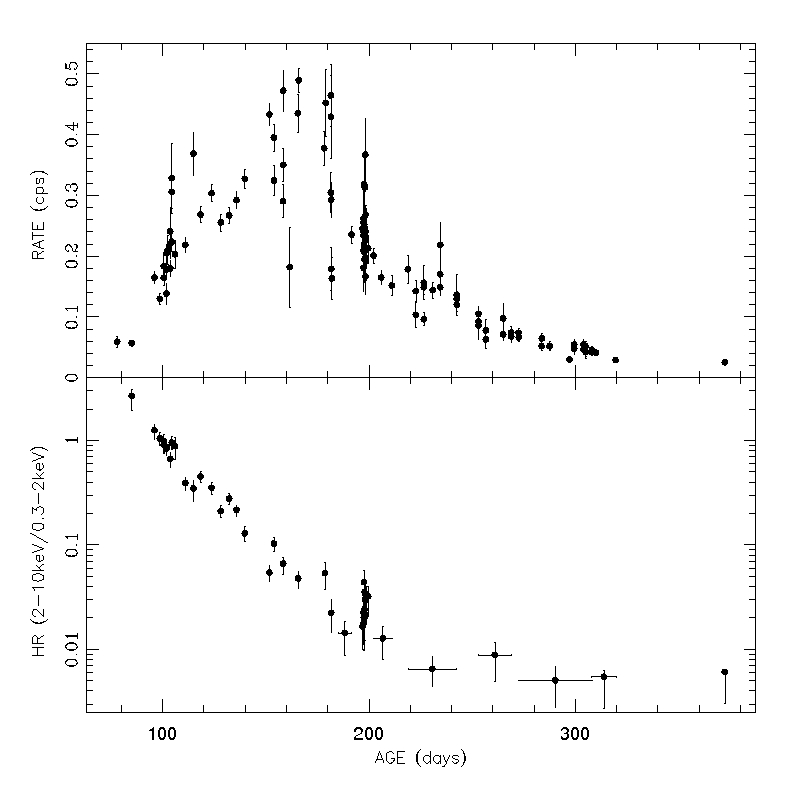}
   \caption{V1369 Cen x-ray (top panel) and hardness ratio (HR, bottom panel) light curves obtained from {\it Swift}-XRT observations. \label{fig:xray}}
   \end{figure}

   \begin{figure*}
   \includegraphics[angle=270,width=18cm]{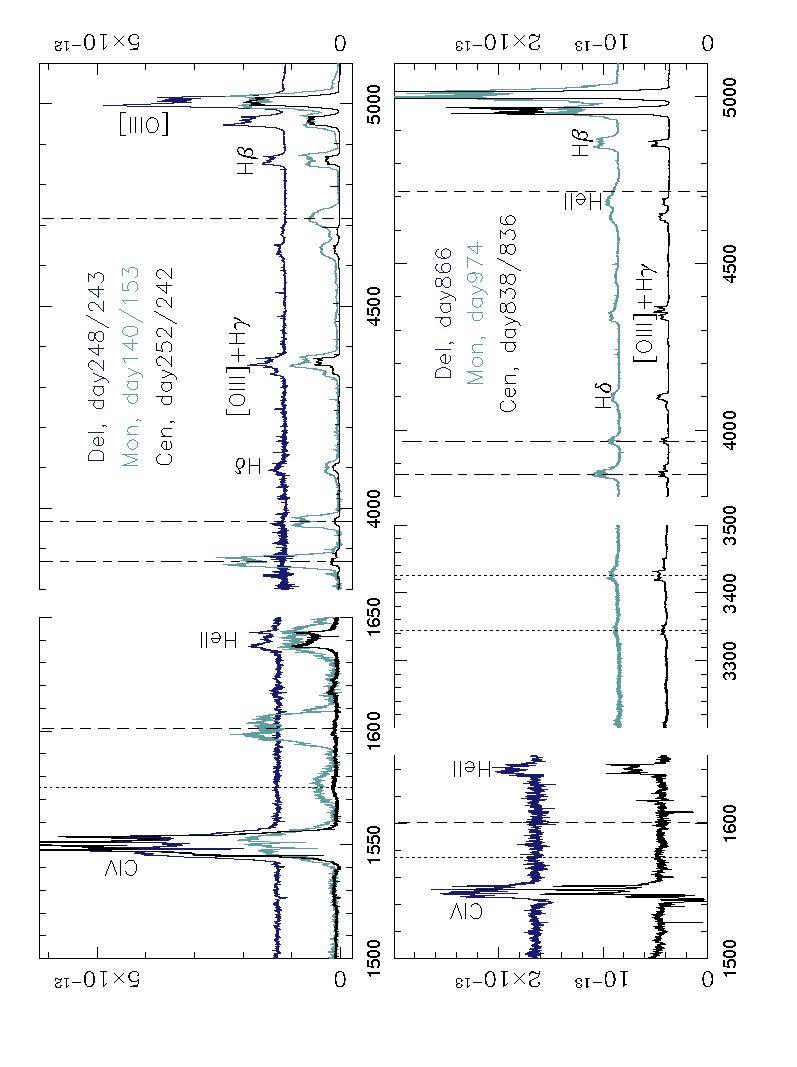}
      \caption{Comparison of V1369 Cen, V339 Del, and V959 Mon Ne emission lines at distinct epochs. Epochs and color-code identification of each spectrum is in the figure itself. The abscissa is in \AA\, and the ordinate in erg/s/cm$^2$/\AA. Note that top and bottom panels cover different wavelength ranges. All spectra, but V1369 Cen optical spectra of days 242 and 835, have been smoothed with a running boxcar filter of size in the range 5 to 19 points, depending on the S/N of the spectrum. The spectra have been scaled and shifted for an easier comparison: in the top panel, V339 Del has been shifted by the constant 1.1$\times10^{-12}$, V959 Mon has been scaled by a factor 10 and V1369 Cen has been scaled by a factor 1/18; in the bottom panel, V339 Del has been shifted by the constant 1.5$\times10^{-13}$, V339 Del has been scaled by the factor 20 and shifted by the constant 0.8$\times10^{-13}$, and V1369 Cen has been scaled by the factor 0.2. No spectrum has been corrected for reddening. The vertical lines 
mark the Ne transitions: the dotted line is for [NeV]\,$\lambda$1575\AA and $\lambda\lambda$3346,3426\AA, 
dashed line is for [NeIV]\,$\lambda$1602\AA and $\lambda$4724\AA, while long and short dashed line is for [NeIII]\,$\lambda\lambda$3869,3968\AA.  \label{fig:NeCOMP}}
   \end{figure*}

   \begin{figure}
   \centering
   \includegraphics[width=9cm]{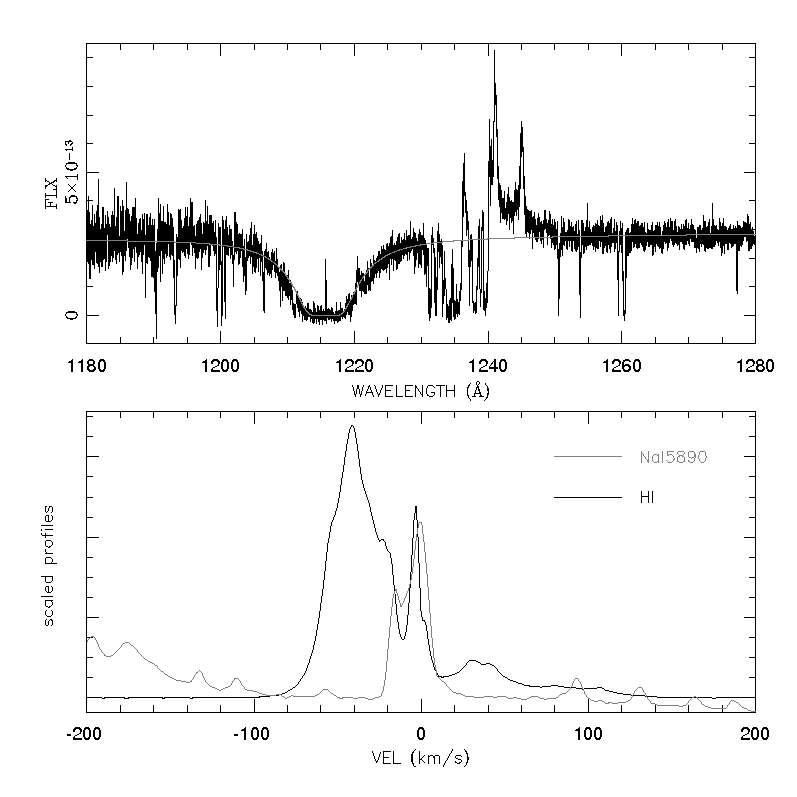}
   \caption{Top panel: Voigt profile fit of the Ly-$\alpha$ interstellar absorption in the 837 day V1369 Cen spectrum. Bottom panel: comparison of the HI line profile from the GASS survey with the NaI interstellar absorption as seen against V1369 Cen's maximum and early decline spectra. Note that the NaI absorption has been inverted for an easier comparison. \label{fig:lya} }
   \end{figure}

\subsection{The x-ray light curve}

As mentioned in the Introduction Section and in Section~\ref{sec:nebularSETs}, the nova x-ray light curve was used to aid the planning the observations and the identification of the nebular spectra. V1369 Cen light curve obtained from Swift observations is shown in Fig.\ref{fig:xray}, together with the hardness ratio evolution.  

The Swift-XRT data were processed and analyzed using the standard HEASoft tools. The optical brightness of the nova was such that the pile-up of optical photons up to x-ray energies (`optical loading') was a concern. Thus, the light curve was obtained extracting only single pixel (grade 0) events, which helps to mitigate this problem. In addition, the earliest
data (up until 2014-02-18; day 78) were obtained using Windowed Timing (WT) mode, which further decreases the optical loading issues, due to the faster read-out time.

At no point was there any sign of x-ray pile-up in the data, so the source light curve and hardness ratio bins were extracted using circular regions with radii of 10-20 pixels (1 pixel = 2.36") depending on the brightness of the source. For the WT data, the background determination was performed using the same size region, offset from, but close to, the source; for Photon Counting (PC) mode, a larger circular region was used.

\section{Selection of the late nebular spectra data sets} \label{sec:nebularSETs}
The reason for concentrating on the late nebular spectra is that they provide essential information about the ejecta structure, kinematics, physic and geometry as we will show. The late nebular spectra are also somewhat easier to deal with than the spectra at earlier stage because they are optically thin. These spectra, however, need to be carefully identified. 
A nova spectrum has usually been called ``nebular'' as soon as [OIII]\,$\lambda\lambda$5007,4958\AA\, appears. Observationally, this occurs at the beginning of the super-soft source (SSS) phase, i.e. when the ejecta are gradually turning optically thin but the density is still high enough that collisions are an important cooling mechanism. Collision and photoionization both affect the ejecta substructures, depending on their positions and densities.  

Here, we define as ``late nebular'' the data collected once the ionization structure of the ejecta has frozen and the ejecta cooling is driven by recombination alone which is mediated by the expansion of the gas (e.g. Vanlandingham et al. 2001). This stage can be identified {\it a priori} from the inspection of the x-ray light curves (total count and hardness ratio) and verified {\it a posteriori} from the inspection of the line profiles (which no longer change) and the development of the integrated line fluxes (which decrease as a power law). 
For each nova we set the start of the late nebular phase on the basis of the visual inspection of the x-ray and hardness ratio (HR) light curves. In this way, late nebular spectra are those obtained after day 250 for the case of V1369 Cen  when the count rate dropped below 0.10 cps and the hardness ratio was less than 0.01 (see Fig.\ref{fig:xray}); after day 280 for V339 Del when the count rate had dropped by almost four order of mag with respect to maximum to $\sim$0.01 cps and the hardness ratio was of the order of 0.1 (see figure 1 of Shore et al. 2016); and after day 250 for V959 Mon when the count rate was about one orders of magnitude below maximum at $\sim$0.1 cps (see figure 1 of Page et al. 2013).  

\section{V1369 Cen spectral characteristics and evolution}
In this and the following section we place V1369 Cen in context determining similar qualitative and quantitative information as in our studies of T~Pyx, V339 Del and V959 Mon (T~Pyx: Shore et al. 2011, 2013a, and De Gennaro Aquino et al. 2014; V339 Del: Shore et al. 2016, V959 Mon: Shore et al. 2013b).  
V1369 Cen offers the most valuable sequence for the study of line profile evolution after the turn off of the SSS, since we have (see Table 1) five epochs between day 252 and 837 and exquisite S/N ($>10$ and $\geq$50 in the UV and optical continuum, respectively, in the last epoch spectra). We note that day 252 was at least $\sim$50 days after x-ray maximum. 

On day 252/242 (where the two numbers indicate the nova age at the time of the UV and optical spectroscopy, respectively) the observed emission lines are those of a highly ionized gas enriched in CNO. The gas emission measure is still fairly large as shown by the line width (FWHM$\sim$1300 km/s) and the extended wings (up to $\pm$1300 km/s from the line center). The line profiles are either ``flat-top'' or ``saddle'' shaped (see Appendix A and Fig.\ref{fig:masters} top panel). The resonant transitions (NV\,$\lambda\lambda$1238,1242\AA, CIV\,$\lambda\lambda$1548,1550\AA, and CII\,$\lambda\lambda$1334,1335\AA) also show a weak and relatively narrow absorption trough in the range $\sim -$1000 to $-$1700 km/s.

The STIS spectrum of day 304 is nearly identical to that of day 252: it displays almost same continuum level (with just a slight change in slope, being steeper in the later epoch) and a small decrease in the lines flux.  

A significant change in the line profiles -especially of the UV range- is observed from day 458/489 spectrum on (see Appendix A and Fig.\ref{fig:masters} bottom panel). The spectra from day 458/489 to day 837/835 show the same dominant species and line profiles, and differ in the line fluxes because of the reduced emission measures. 
The emission component is confined to narrow peaks at velocity $-$600 and $+$500 km/s, and the resonance lines that showed narrow absorption troughs now display multiple discrete absorptions in the range $\sim-$700  to $-$2000 km/s. Forbidden and recombination transitions display similar emission peaks but they also show substantial emission at low velocities in the range  $-$600$<$v$_{rad}<+$500 km/s. The peaks appear more pronounced in the higher excitation transitions and in the  [NeIII]\,$\lambda\lambda$3869,3968\AA\, and [NII]\,$\lambda$5755\AA\,$\lambda\lambda$6583,6548\AA\, lines. 
The weak UV lines have decreased in flux so the spectrum is now dominated by NV\,$\lambda\lambda$1548,1550\AA, CII\,$\lambda\lambda$1334,1335\AA, OIV\,$\lambda\lambda$1401,1405\AA, NIV]\,$\lambda\lambda$1483,1486\AA, CIV\,$\lambda\lambda$1548,1550\AA, HeII\,$\lambda$1640\AA, and CIII]\,$\lambda\lambda$1906,1908\AA. In the optical, thanks to the larger wavelength coverage of the UVES spectrum, we detect OIII lines pumped by the Bowen fluorescence mechanism (e.g. $\lambda$3132\AA, \,$\lambda$3340\AA, $\lambda$3444\AA, and $\lambda$3759\AA), and [NeV]\,$\lambda\lambda$3346,3426\AA. The optical lines have also weakened, but emission from [ArV]\,$\lambda$7006\AA\, and [FeVII] (e,g. $\lambda$3586\AA, $\lambda$5158\AA\ and $\lambda$5721\AA) are now relatively stronger. 

One aspect worth stressing for V1369 Cen is the observed set of neon lines and their relative strengths compared with the other novae spectra. V959 Mon was classified as an ONe nova on the basis of its comparison (and very close match) with the nova Vel 1999, nova Cyg 1992 and nova LMC 2000 spectra at similar ages (Shore et al. 2013b), and V339 Del was classified as a CO nova by comparison with OS And (Shore et al. 2016). V1369 Cen does not resemble any of those objects. Its Ne line strengths are atypical for an ONe nova, since it never develops strong UV Ne transitions, while [NeIII] and [NV] lines are quite strong and, in particular, much stronger than in CO novae at any stage (see Fig.\ref{fig:NeCOMP}). Moderate Ne enrichment is expected in CO novae since during the He burning phase $^{14}$N is converted in $^{22}$Ne, or in case the nova experienced the CNO breakout during the thermonuclear runaway (TNR, e.g. Livio \& Truran 1994, Jos\'e \& Shore 2008). V1369 Cen is likely one of the moderately Ne 
enriched nova as defined by Livio \& Truran (1994). 

\section{V1369 Cen physical parameters}
\subsection{Reddening}
 
From a series of 12 spectra (FEROS, R=48000) taken during early decline, we measured an equivalent width EW=0.345$\pm$0.005 \AA\, for the NaI D2 5890 \AA \ interstellar absorption, corresponding to E(B-V)=0.15 mag according to Munari and Zwitter (1997) method. Similarly, measuring the EW of the separate components and adding the individual E(B-V) we obtain E(B-V)=0.16 mag. However, these E(B-V) may be only lower limits, since the interstellar neutral sodium could have condensed into dust and be depleted in the gas phase. Therefore, for our reddening estimate we also turned to the HI column density. This can be derived by modeling the interstellar Ly$\alpha$ absorption in the day 837 spectrum\footnote{The comparison with previous epoch spectra guaranteed that the Ly$\alpha$ absorption was purely of interstellar origin.} by assuming different column densities (Fig.\ref{fig:lya}, top panel). The best Ly$\alpha$ match is obtained for HI column densities of $\sim$7.2$\times10^{21}$ cm$^{-2}$ corresponding to E(B-V)=0.15 mag, in agreement with the NaID estimate. We additionally checked that the derived column density is consistent with that computed by integrating the HI 21~cm emission in the LAB and GASS maps (Kalberla et al. 2005, Kalberla \& Haud 2015) in the velocity range corresponding to the NaID interstellar absorption profile\footnote{As a consistency check the UV resonance lines OI\,$\lambda$1302\AA, SiII\,$\lambda$1260\AA\, and $\lambda$1526\AA, AlII\,$\lambda$1670\AA\, and CI\,$\lambda$1561\AA\, were also measured.} (Fig.\ref{fig:lya}, bottom panel).  We adopt E(B-V)=0.15 mag and Cardelli et al. (1989) extinction curve when dereddening our data in the following analysis and measurements. 

   \begin{figure}
   \centering
   \includegraphics[width=9cm]{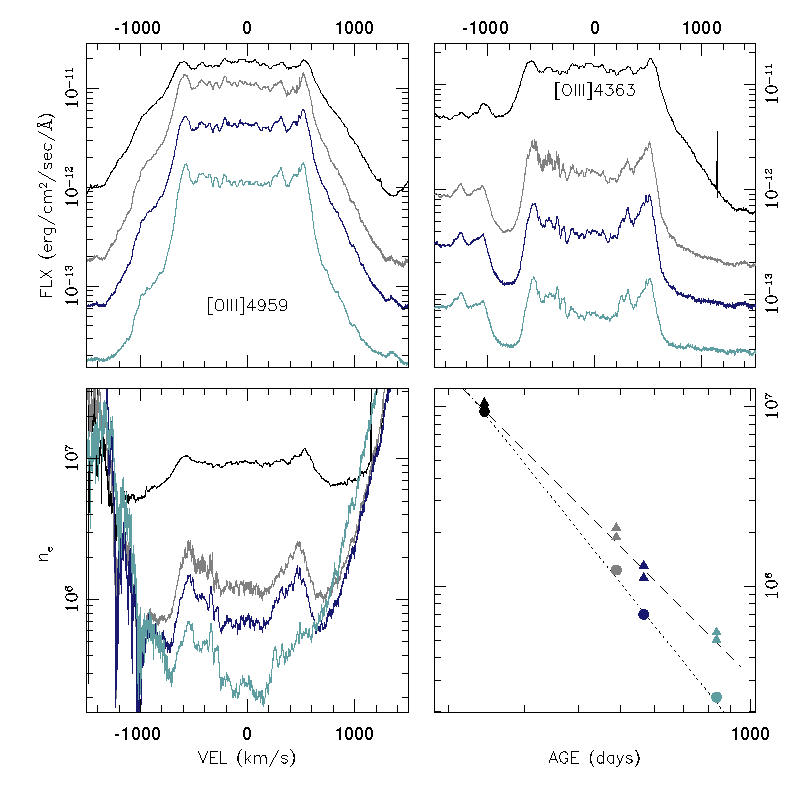}
   \caption{V1369 Cen [OIII] line profiles (top panels) and electron density as derived from the [OIII] emissions diagnostic: in velocity space at various epochs (bottom left panel) and as function of time (bottom right panel). Note that both fluxes (top two panels) and densities (bottom two panels) are in log scale to enhance the contrasts. Different color indicates different epoch as it can be evinced from the bottom right panel. In the bottom right panel solid circles indicate the average density as measured in the velocity interval $-$100 to 100 km/s, while the triangles indicates the average density at the two peaks centered at $-$600 and 500 km/s. 
 \label{fig:cen_ne}}
  \end{figure}

\subsection{Electron density}
We computed the electron density, $n_e$, and derived its variation over time. Despite the optimal wavelength coverage of our spectra, we are often impeded in the use of the classic diagnostics because of the heavy blends or small gaps in spectral coverage. We could not use the [NeV] and [NeIII] because we do not observe the [NeV]\,$\lambda$2975~\AA\, emission, and for [NeIII] we are unable to deblend the [NeIII]\,$\lambda$3342\AA\, from the \,$\lambda$3346\AA\, due to  superposed emission from OIII\,$\lambda$3341\AA\, (all epochs). We could not use the [NII] diagnostic even when the $\lambda\lambda$6548,6583\AA\, doublet was separable from H$\alpha$ because the [NII]\,$\lambda$5755\AA\, emission line falls in the gap of UVES red CCD mosaic in the VLT observations. Instead, we used the [OIII] diagnostic, deblending the \,$\lambda$4363\AA\ from H$\gamma$ line (by scaling H$\beta$ to the blue wing of H$\gamma$ as in our previous studies). 

Unlike previous analysis in the literature, our method does not use the integrated line flux. We computed the electron density for each velocity resolution element. The advantage is that of retaining information about the ejecta projected geometry and kinematics and revealing different local conditions.
We plot in Fig.\ref{fig:cen_ne} the electron density in velocity space at the four epochs of Table \ref{tab:log_cen}, as derived from the [OIII] emission lines. The plot shows how the electron density decreases from $\sim$10$^7$ cm$^{-3}$ to less than 10$^{6}$ cm$^{-3}$ between day 242 and 835, at least for the low velocity ($|$v$_{rad}|\leq$800 km/s) portion of the ejecta. The high velocity gas (i.e. $|$v$_{rad}|\geq$800 km/s) in [OIII] is too weak to allow density computations. 
The low velocity gas, however, is not uniform in density and the portions between -700$\leq$v$_{rad}\leq$-500 and 400$\leq$v$_{rad}\leq$600 km/s have somewhat higher density than the line ``center'' (-500$<$v$_{rad}<$400 km/s). The contrast increases from day 242 to day 835 up to a factor of 2. The innermost portion of the ejecta follows the density decrease expected for ballistic expansion (i.e. $n_e\propto t^{-3}$), while the intermediate portion of the ejecta follows a less steep decrease in time ($n_e\propto t^{-2.4}$, see bottom-right panel of Fig.\ref{fig:cen_ne}). This does not imply that the ejecta is following different expansion laws in different portions, but that the intermediate velocity portion of the ejecta is not yet in the collisionless limit.

  \begin{figure}
   \centering
   \includegraphics[width=9cm]{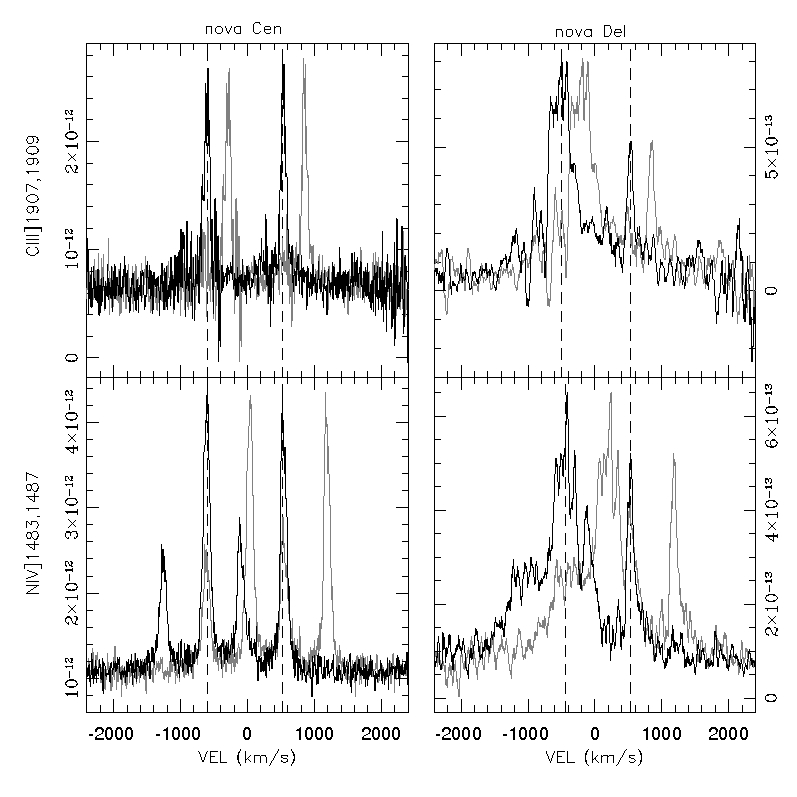}
      \caption{V1369 Cen (left) and V339 Del (right) CIII (top) and NIV (bottom) line profiles at their last epoch. The black color is for CIII]\,$\lambda$1908\AA\ (NIV]\,$\lambda$1486\AA), while the gray is for CIII]\,$\lambda$1906\AA\, (NIV]\,$\lambda$1483\AA). Vertical dashed lines mark the blue and red peaks of the emission line. V339 Del asymmetric profiles indicate geometrical (at the substructure level) asymmetry. Density inhomogeneity is also possible but cannot be verified from the lines themselves because the doublets are blending. \label{fig:ne2}}
  \end{figure}

In addition, the profile of the UV transitions in the last epoch (day 837) has evolved in such a way that the two doublets CIII]\,$\lambda\lambda$1906,1908\AA\ and NIV]\,$\lambda\lambda$1483,1486\AA\ are no longer blended and also can be used as density diagnostics. This is a first for novae, although the diagnostic is standard for HII regions, planetary nebulae and symbiotic stars. {\it Note that these ratios are independent of reddening.} Fig.\ref{fig:ne2} shows (left panels) V1369 Cen CIII] and NIV] line profiles and, for comparison (right panels) the same transitions in the last spectrum of V339 Del (day 866). The doublets are blended in V339 Del, even at the last epoch, and the line profile shows an obvious asymmetry of the ejecta. We will return to this in Section 6 when discussing the geometry.  

Nussbaumer \& Schild (1981) plot the NIV] line ratio for pure recombination as a function of the electron density for a range of temperatures. We measured the NIV] line ratio for the blue and red peak separately and ascribe any difference to noise.  We obtain an average line ratio of 0.52$\pm$0.06 so that $n_e\simeq$5.2$\times10^5$ cm$^{-3}$, compared with the average value of 5.6$\times10^5$ cm$^{-3}$ from the [OIII] density diagnostic at the blue and red peak. Note that the [OIII] and NIV] peaks yield both comparable densities and density decline rate (as computed for the NIV from the last two epochs). 
The C$^{2+}$ recombination coefficient and line ratio as function of density have been computed by Nussbaumer \& Schild (1979). We measured the average flux ratio 0.30$\pm$0.05 from the blue and red peak of the CIII] transition in the day 837 spectrum (after a boxcar smoothing of 15 resolution elements).  Assuming the electron temperature T$_e$=10000 K we derive from Nussbaumer \& Schild (their figure 5a) $n_e\sim$1.4$\times10^5$ cm$^{-3}$. 

\subsection{Abundance ratios from absorption features}

\begin{figure}
   \centering
   \includegraphics[width=9cm]{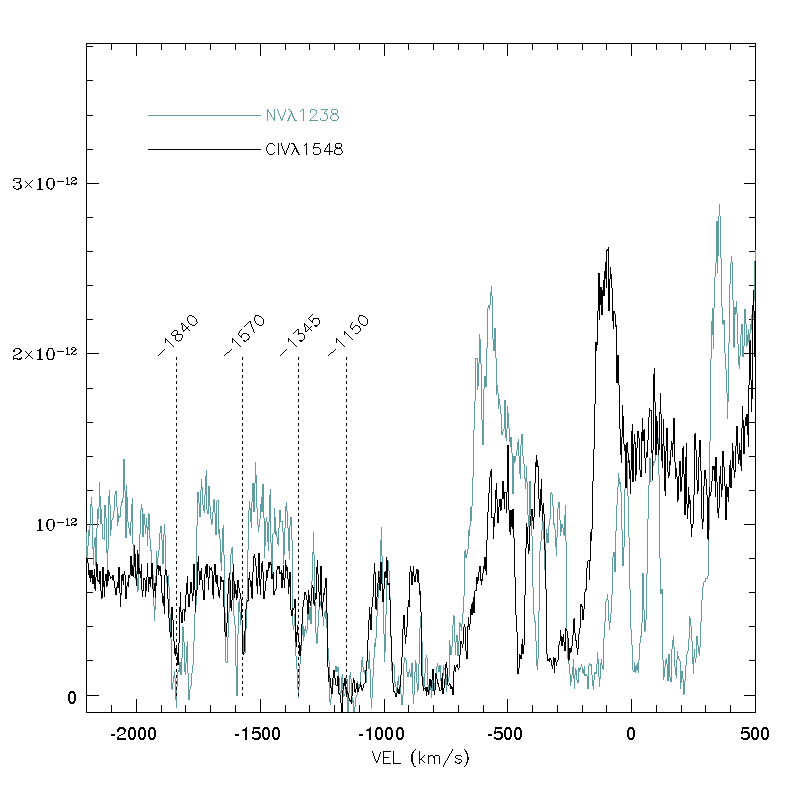}
      \caption{The absorption structures observed in the day 837 spectrum for the transitions NV$\lambda\lambda$1238,1242\AA\, and CIV\,$\lambda\lambda$1548,1550\AA. Vertical dotted lines mark the approximate wavelength of the absorption components used for the N/C ratio computation. The y-axis is flux in ergs/cm$^2$/s/\AA.  
         \label{fig:ncr}}
\end{figure}

A striking important result can be obtained from the comparison of the discrete absorption features for NV\,$\lambda\lambda$1238,1242\AA\, and CIV\,$\lambda\lambda$1548,1550\AA. In V1369 Cen we observe a significant number of coincident (in velocity) NV and CIV absorptions, whose equivalent width, EW, were used to obtain the N$^{4+}$/C$^{3+}$ column density ratio as follows: 
\\

\noindent (N/C)$_{v_{rad}}$=(EW$_N$/EW$_C$)($\lambda_C/\lambda_N$)$^2$($f_C/f_N$)$\sim1.8$(EW$_N$/EW$_C$)
\\

\noindent where $\lambda$ is the rest wavelength and $f$ is the oscillator strength. The common velocity components were at $-$1840, $-$1570, $-$1345 and $-1150$ km/s (see Fig.\ref{fig:ncr}), and the derived N$^{4+}$/C$^{3+}$column density ratio was 270, 200, 233 and $>$120, respectively, with an uncertainty of about 10\%. Thus, excluding the lower limit and assuming that the measured resonance lines represent the dominant ion state, the ratio of nitrogen to carbon was $\geq$200. The solar ratio is 0.31 (Asplund et al. 2009) implying that V1369 Cen was enormously enhanced in nitrogen relative to carbon. The ratio is more extreme than published nucleosynthetic yields of the TNR of massive WDs (e.g. Downen et al. 2013, Starrfield et al. 2016, Casanova et al. 2016). 

A similar analysis applied to T~Pyx and V339 Del produced different values. In particular, we found N/C ratio in the range [1,6] for the case of V339 Del\footnote{Note that the value is quite uncertain, since V339 Del absorption features remained less distinct even in the last epoch.}, and N/C$\simeq$2.3$\pm$0.2 for T~Pyx. In both cases, the values are not as different relative to the solar value as for V1369 Cen, and are compatible with theoretical TNR predictions. 

Note that in all cases no information is available for the isotopic composition of the clumps. Although consistent with the general expectation for the explosive yields, these abundance ratio cannot place precise constraints on the thermodynamic condition at the peak of the TNR. 

\subsection{Filling factor \& distance}
The first STIS spectra were taken at the highest echelle resolution when the nova was especially bright.  
This affords the best interstellar absorption line profiles and we have used the OI\,$\lambda$1302\AA, SiII\,$\lambda$1260\AA, CI\,$\lambda$1561\AA, and AlII\,$\lambda$1670\AA\ lines to provide a range of ionizations and line strengths. Except for OI, all are strong but unsaturated.  The interval covered by the profiles was $-11 \le$ v$_{rad,helio} \le 14$ km/s with mean velocity of 1.4 km/s (weighted flux distribution), corresponding to v$_{rad,LSR} = -$2.4 km/s in the local standard of rest (LSR).  Using the IAU standard Oort constants and distance to the Galactic center, this velocity implies a distance $1.8 \le D(kpc) \le 2.4$.  We adopt 2 kpc in what follows.  This is also consistent with the extinction.  The distance is independent of any Maximum Magnitude versus Rate of Decline (MMRD) assumptions.  As we discussed, V1369 Cen did not resemble either V959 Mon (or other ONe classical novae) or V339 Del so we cannot scale the spectra relative to previously observed exemplars. 

For mass determination, we take advantage of the multiple determinations of the $n_e$ from the last epoch optical and UV spectrum (day 835, 837).  The measured H$\beta$ flux was $F(H\beta) = 1.47\times 10^{-12}$ erg s$^{-1}$ cm$^{-2}$.  With a reddening E(B-V)=0.15 applied and for a distance of 2 kpc, the line luminosity was $L(H\beta) =  1.2\times 10^{32}(D/2\ kpc)^2$ erg s$^{-1}$. The range of $n_e$ obtained from the CIII]\,$\lambda$1908\AA, NIV]\,$\lambda$1486\AA, and [OIII] nebular lines is $1 <$ (n$_e/10^5$) cm$^{-3} < 6$ giving a predicted range $0.5 < L(H\beta)/(10^{35}$\ erg\ $s^{-1}) < 12$ with a choice of v$_{max}$=2500 km s$^{-1}$.  Although this is a broad range, other quantities are fairly well constrained: the solid angle of the aspherical ejecta as provided by the bipolar models (see Section \ref{sec:sec_mc}) is $(\Delta \Omega /4\pi) \approx 1/3$  and the filling factor is in the range of $0.1 < f < 0.2$. Combined with the electron density, the ejecta mass resulted approximately $(1.1\pm 0.4)10^{-4}$M$_\odot$  adopting a solar He/H ratio.  Although some novae appear to have either hydrogen deficiencies or enhanced helium abundances (e.g. Jos\'e \& Shore 2008; Schwarz 2014), the electron density is obtained directly from the plasma diagnostics and not from the H$\beta$ flux, which was used only to derive the filling factor. The filling factor scales as $f\sim [1+n(He)/n(H)]^{-1}$ and the mass scales as $(\mu/1.6)f$.  The independent determination of the helium abundance is only possible with detailed modeling of the recombination and ionization structure of the ejecta and applies only during the photoionization dominated stage before SSS turnoff. 
We note, alternatively, that the mass scales as $M_{ej} \sim$ v$_{max}^{-3/2} D n_e [E(B-V)]^{1/2}$.

This mass is higher than we found for V339 Del (2-3$\times$10$^{-5}$\,M$\odot$, Shore et. al. 2016) and V959 Mon ($\lesssim$6$\times$10$^{-5}$\,M$\odot$, Shore et al. 2013b), but not excessively so, and around the values expected from the models of the TNR (Downen et al. 2013).

\section{Ejecta geometry and ionization structure }
In this section we analyze the ejecta ionization structure and infer the ejecta geometry. Note that the ionization structure is a key component in our work since it serves to constrain a self-consistent scenario. 

There are two ways of looking at the ejecta structure in general, and their ionization structure in particular: 1) the emission lines and 2) the absorption components. We explain these in turn.

\subsection{Analysis of the emission lines}\label{sec:emanalysis}

One must never forget that the emission line profile is the result of a line of sight velocity projection effect. For optically thin gas, we see everything that has a sufficient emission measure in a given radial velocity bin, so we measure the sum of the contributions from individual clumps, each of which may have different density conditions. Hence, even the density in velocity space, as in Fig.\ref{fig:cen_ne}, is an average per velocity bin.  The diverse local conditions of the clumps imply that different parts of the ejecta might develop differently  depending on their different outward velocity relative to the WD. However, once each clump is in pure recombination for a given ion, the line profile is ``frozen'' and the line flux simply decreases with time with no change in v$_{rad}$. With this in mind, we can compare lines produced by different mechanisms (e.g. recombination, collisions exciting a metastable level or intercombination transitions) and by different ions of a same element. 
These comparisons yield the boundaries of the distribution of various ions, i.e. the ionization structure, and help disentangling abundance from ionization effects in the observed line profiles. 

To illustrate this for the V1369 Cen ejecta we compare: 1) the temporal evolution of three forbidden transitions of oxygen from different ionization stages (O$^o$ to  O$^{2+}$); 2) the different profiles of recombination, intercombination and forbidden transitions from low and high ionization energy ions at epoch $\sim$566 days; 3) the profiles of recombination lines from different ions at day 835. 

\begin{figure}
   \centering
   \includegraphics[width=9cm]{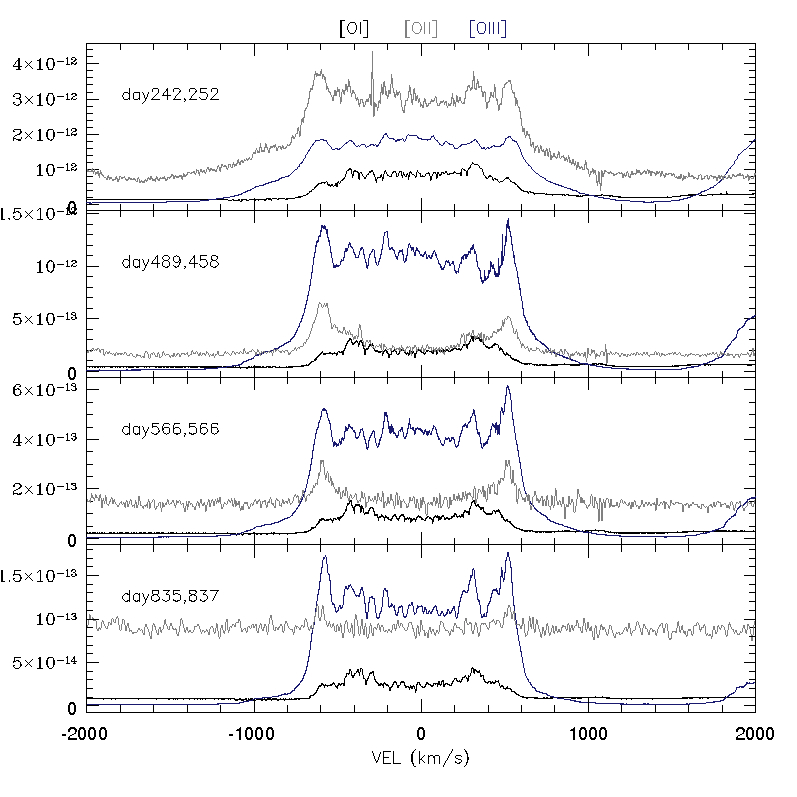}
      \caption{V1369 Cen ionization structure as evinced from the nebular lines [OIII]$\lambda$4959\AA, [OII]$\lambda$2470\AA\, and [OI]$\lambda$6300\AA\,  at the four epochs. Note that [OIII]$\lambda$4959\AA\ has been scaled by a factor 0.1, while [OII]$\lambda$2470\AA\ has been scaled by a factor 0.4.  
         \label{fig:ionizationStrct}}
\end{figure}

  \begin{figure*}
   \centering
   \plottwo{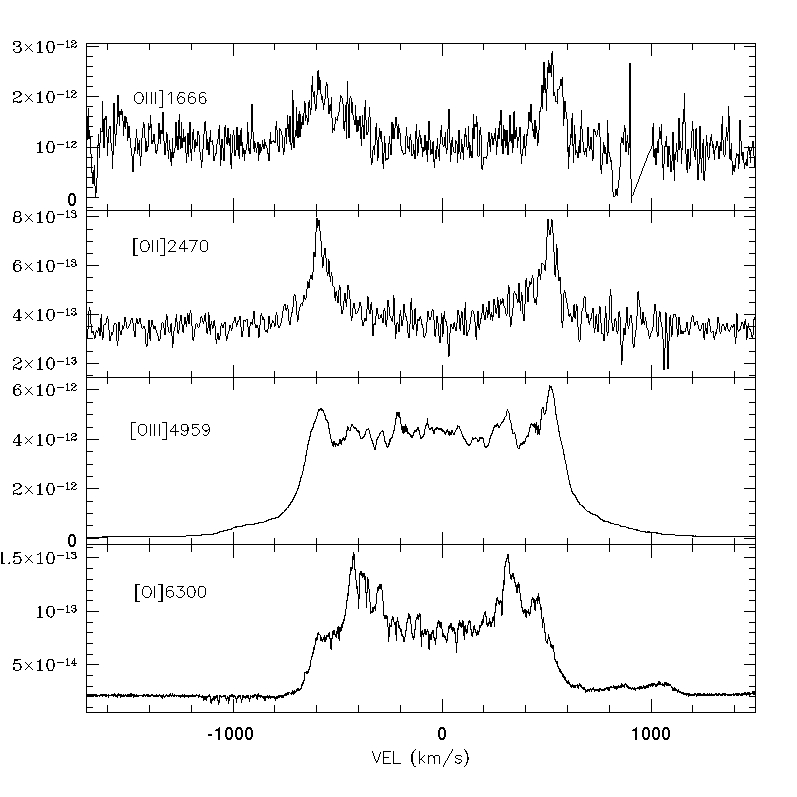}{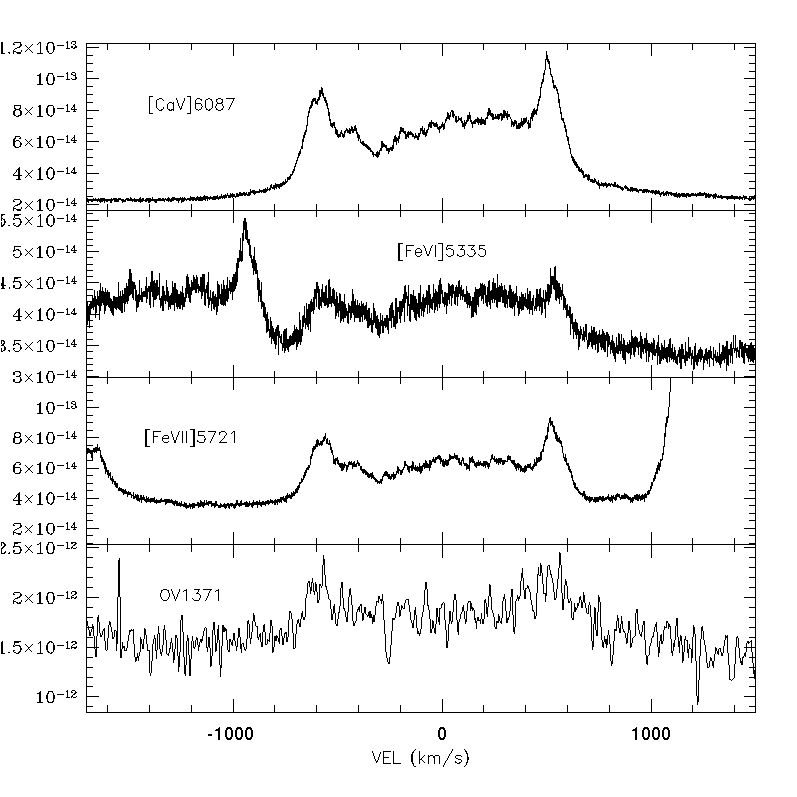}
   \caption{Line profiles of different ions in V1369 Cen spectrum of day 566. Left panel: intercombination and forbidden transition from low energy ions of oxygen (the ionization potential (IP) of O$^{4+}$ is $\sim$55eV). Right panel: forbidden and recombination transitions from high energy ions: IP(Ca$^{4+}$)$\simeq$67ev, IP(Fe$^{5+}$)$\simeq$76eV, IP(Fe$^{6+}$)$\simeq$100eV, IP(O$^{6+}$)$\sim$113eV.
         \label{fig:Osnapshot}}
   \end{figure*}
   
\begin{figure}
   \centering
   \includegraphics[width=9cm]{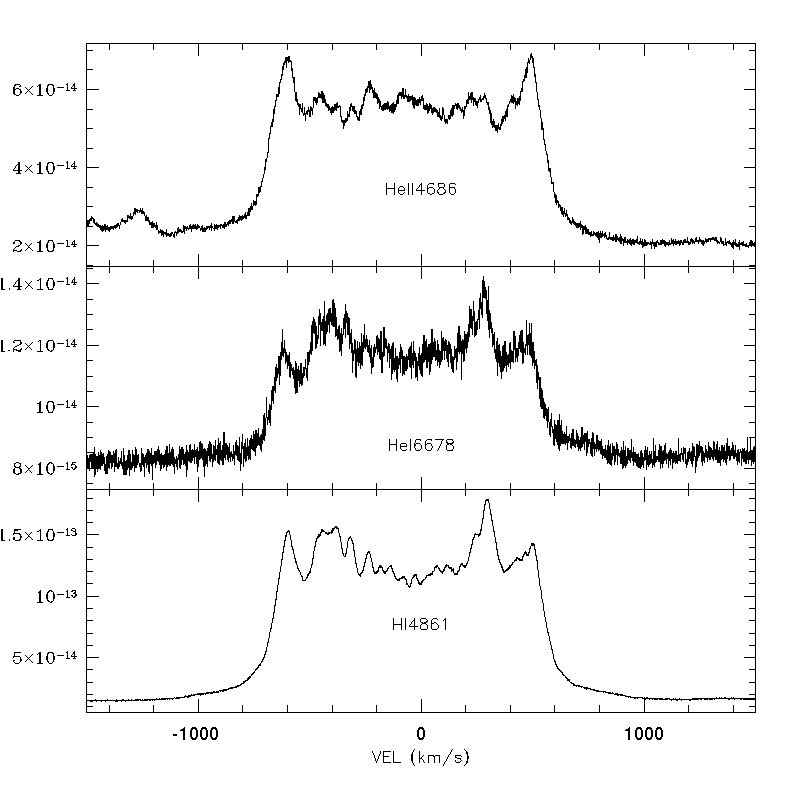}
   \caption{V1369 Cen H and He recombination line on day 835. On the y-axis of the plot, fluxes are in erg/cm$^2$/s/\AA.
         \label{fig:rec}}
   \end{figure}

The evolution of the oxygen ions is shown in Fig.\ref{fig:ionizationStrct}. Since this treats a single element in various ionization stages, it shows the development of the ionization structure unbiased by abundance. It is interesting to note here that in the first epoch (day 242/252), [OII] and [OIII] both have very pronounced wings (800$<|$v$_{rad}|<$1600 km/s) which quickly disappear for the [OII] transitions but persist (although weakened) in [OIII]. Hence, in that velocity range the densities are low enough that O$^{2+}$ no longer recombines to O$^{+}$ as easily. Oxygen exists mostly in the double ionized state which is excited from the ground level by collisions (and the collisional excitation cross section is larger than the collisional recombination cross section). Similarly the neutral O is confined to higher density regions that correspond to the lower velocity bins. By projection, the peak intensity of the [OI] transition occurs at $-$400 and +300 km/s while the [OIII] emissions have their peak intensity at $-$600 and +550 km/s\footnote{Note that also recombination lines display peaks at those same velocities}. At those velocities the [OI] shows only a weak shoulder. This implies that the O$^{2+}$ concentrates, on average, at larger velocities with respect to O$^0$, i.e. the O$^{2+}$/O$^o$ ratio increases at higher velocity. The lowest v$_{rad}$ portion of the emission lines (velocity range $-$300$<$v$_{rad}<$200) shows relatively higher flux in the [OIII] transition than in the [OII] and [OI], especially at epochs 458 and 566, indicating that there are projected low velocity regions where O$^{2+}$ does not recombine. In other words, the oxygen is mostly in the O$^{2+}$ state everywhere across the ejecta but is relatively more abundant --compared to O$^+$ and O$^0$-- in the outer portion of the ejecta (those with the higher velocity) where the density must be lower for the recombination rate to be consequently lower. 
The O$^+$ tracer, the [OII]$\lambda$2470\AA, disappears earlier  (at day 837 it is almost completely gone) because of the higher energy of its upper level.  

Fig.\ref{fig:Osnapshot} shows forbidden, intercombination and recombination transitions from different ions observed on day 566. In particular, the left panel compares line profiles from low energy oxygen ions including the OIII]$\lambda$1666\AA\ line that is an intercombination transition and shows where O$^{3+}$ recombines to O$^{2+}$.  
The transition shows emission at the -600 and +500 km/s peaks and is undetectable everywhere else, i.e. the emission measure is larger where the density is not too low to prevent the O$^{3+}$ recombination and not too high to favor the recombination of lower ions. 
The right panel compares forbidden and recombination transitions from high energy ions (67eV$\leq$IP$\leq$113eV, where IP is the ionization potential), showing that their profiles are similar, i.e. the ions are co-spatial. We note that the OV$\lambda$1371\AA\, line and similar recombination transitions (e.g. NIV$\lambda$1718\AA) are frozen since high energy ionizing photons are no longer dominant. The highest energy ions provide information about the spectral energy distribution of the underlying photoionizing source: the detection of OV and [FeVII] lines and the lack of [FeX] constrain the supersoft source (when on) to temperatures $>$100-130 eV and $<$200 eV. 

Finally, we show in Fig.\ref{fig:rec} the profiles of the recombination transitions HeII$\lambda$4686\AA\ and the singlet HeI$\lambda$6678\AA\ together with H$\beta\,\lambda$4861\AA\, at day 835. The two He lines show the distribution of singly and doubly ionized He; since H is the most abundant element, H$\beta$ shows the projected geometry of the ejecta. The HeI profile resembles the [OI] line but with more pronounced peaks at $-$400 and +300 km/s. In contrast, HeII has peaks at $-$600 and +500 km/s that are very similar to [OIII], although the two HeII peaks have greater contrast relative to the rest of the line than those in the [OIII] line profile. The similarity is not surprising, since HeII and [OIII] trace the distribution of the highly ionized gas, while HeI and [OI] trace the neutral gas. The individual different contrasts in each line result from the fact that recombination lines have much higher transition probability than forbidden lines and these also depend on their collisional cross section and the electron energy distribution for the population of their upper level. 

\subsection{Monte Carlo simulation of the emission lines} \label{sec:sec_mc}

\begin{figure}
   \centering
   \includegraphics[width=9cm]{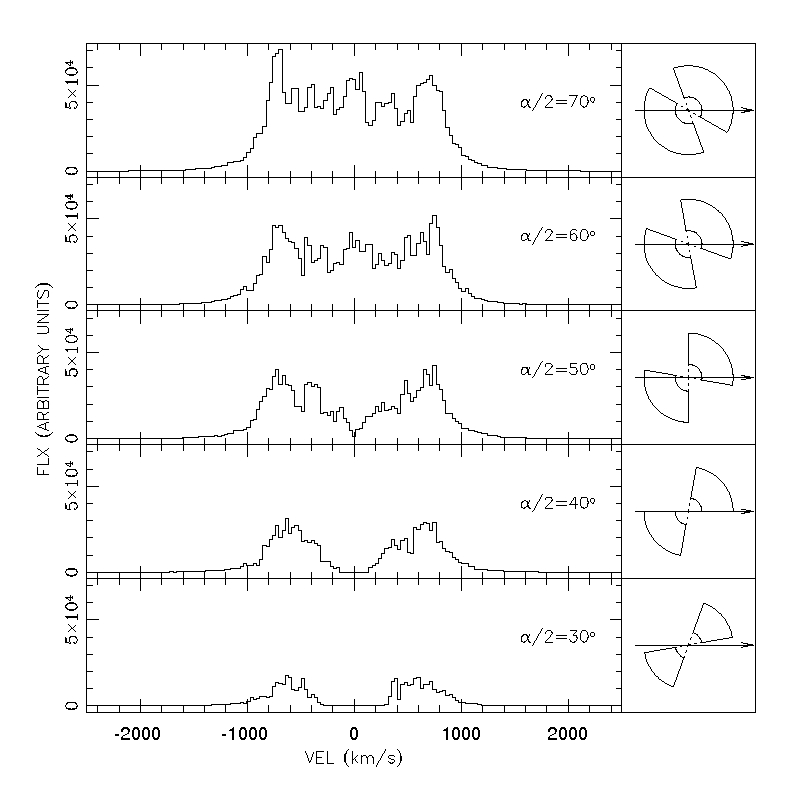}
   \caption{Line profiles produced by a biconical ejecta as simulated by Shore et al. (2013a). In all the cases it is assumed 1) an inclination of 40 deg of the observer with respect to the bi-cone axis, 2) an ejecta thickness of 0.7$\times R$ where, $R$ is the maximum ejecta radius, and that the truncated cone is filled in. The simulation differs in the opening angle $\alpha$ of the cone, indicated in the figure itself. A 2D sketch of the corresponding ejecta geometry is indicated on the right of each profile. The arrow points to the observer. 
         \label{fig:mod_e}}
   \end{figure}

In this section we apply the Monte Carlo simulation already used for T Pyx, V959 Mon and V339 Del, to reproduce the observed line profiles in the V1369 Cen spectra. The code uses random particles confined in a biconical geometry to reproduce the gross characteristics of the line profiles. The random particles have linear velocity law within a range matching the observations, constant mass, and emissivity corresponding to the evolutionary stage of the nova (recombination). Given the  frozen state of the ejecta, it would be physically inconsistent to adopt any photoionization code to reproduce  the observed line profile. Photoionization codes assume equilibrium (which is not physically appropriate) and do not account for expansion. 
The simulation reproduces the global line profile but {\it not} specific substructures: these result from the simulation noise, but correspond to individual clumps overlapping in the projected velocity space, in real ejecta. 
Modeling of individual substructures would require adopting specific particle distributions, an assumption that is not relevant to the purpose of our work.

\begin{figure}
   \includegraphics[width=9cm]{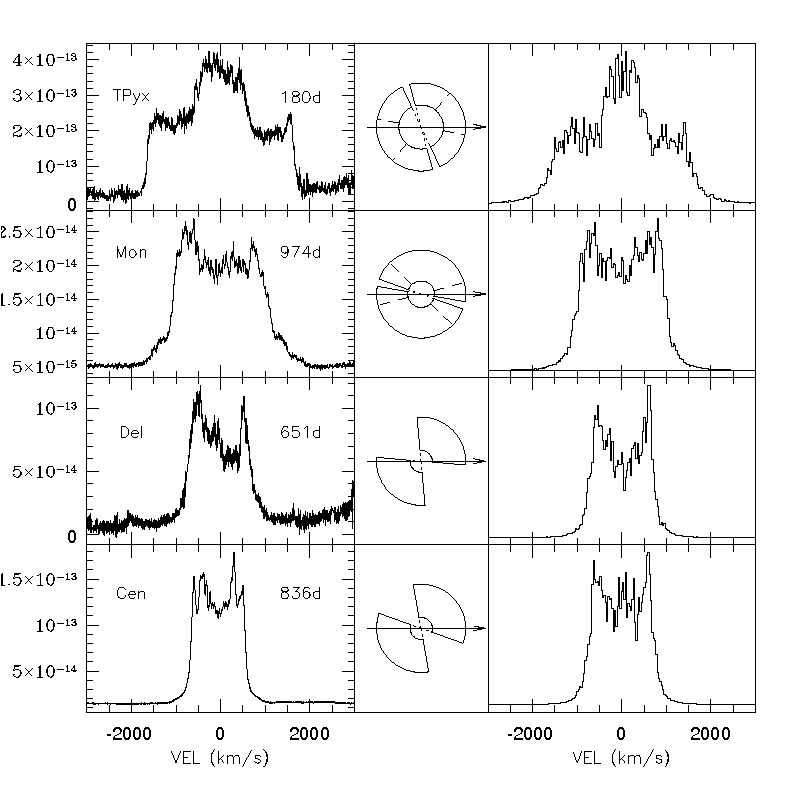}
   \caption{Left column: H$\beta$ emission line profiles at the last epoch of observation for each of the ``panchromatic novae'' discussed in this series of publications. The ordinate is flux in units of erg/cm$^2$/s/\AA. The identification of each nova is on the plot itself together with the age of the nova (i.e. the epoch of observation in days). T~Pyx and V959 Mon spectra have been smoothed by a running boxcar of 5 points. Middle and right column: the biconical geometry derived for each nova and the corresponding predicted line profile. Dashed lines in the model geometry from T Pyx and V959 Mon indicate a hollow cone (see Shore et al. 2013a and Shore et al. 2013b for more details). 
         \label{fig:hb}}
\end{figure}

A variety of profiles can result from our code, including the most common, so called, ``saddle'' and ``castellated'' shaped profiles. They are produced by a combination of opening angles, cone thickness and inclination with respect to the line of sight. Fig.\ref{fig:mod_e} shows the effect of the cone opening angle in shaping the line profile for a fixed inclination. Less collimated ejecta produce more ``rectangular'' (or ``flat-top'') profiles, while a ``saddle'' or v-shaped profile are obtained with more collimated ejecta.
In the case of V1369 Cen we could reproduce its ``horned'' line profiles of width $\sim$1200-1500 km/s adopting a biconical geometry with opening angle up to 140$^o$, thickness 0.75$R$ (where $R$ is the maximum radius of the ejecta) and maximum expansion velocity v=2500 km/s viewed at an inclination of $\sim$40$^o$. 
Fig.\ref{fig:hb} displays the H$\beta$ profile for T Pyx, V959 Mon, V339 Del and V1369 Cen, together with the biconical geometry that best mimics their profile and the corresponding Monte Carlo simulation predicted profile.
The figure shows the gross similarities and the specific differences characteristic to each ejecta. The emission lines of a given nova can be broader or narrow and more or less rectangular in shape. They can also display extended wings as in the case of T~Pyx. All trace back to the specific characteristic of the biconical ballistic ejecta. 

Note that the simulations are not fitting each profile but providing a tool for interpreting the data. In particular: 1) for each nova, the opening angle varies with the transition as different ions have different structures (e.g. [OIII] typically suggests a slightly wider opening angle than [OI] given that the most external regions are less dense and, therefore, more highly ionized, see Section \ref{sec:emanalysis}); 2) there might be degeneracies in the simulation (see Shore et al. 2013a); 3) the line widths and the emission measure decreases with time; 4) the simulations produce globally symmetric profiles. Asymmetries such as those observed in V339 Del at each epoch and in all transitions (see Shore et al. 2016) likely reflect intrinsic characteristics of the ejecta whose origins remain to be investigated.  Hence, the take away result from the simulation is that all of the gross features of the observed profiles for each nova can be reproduced by a biconical geometry with relatively wide opening angles and relatively large thickness. Features resembling rings, equatorial belts, and polar caps also result from projection effects of the biconical geometry onto the sky plane (e.g. see figure 10 in Shore et al. 2013a and figures 16 and 20 in Shore et al. 2013b). Thus, bipolar ejection seems to be common to all novae without requiring additional specially tailored  structures such as equatorial rings or jets.

\subsection{Analysis of the absorption features}

\begin{figure*}
   \plotone{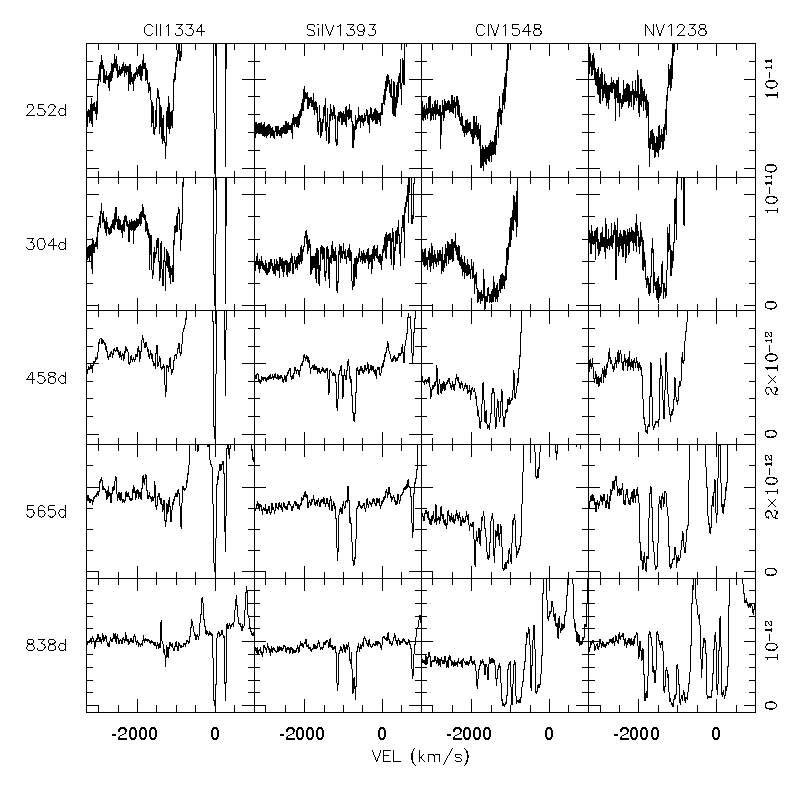}
   \caption{V1369 Cen ionization structure as evinced from the absorptions observed in the permitted resonant transitions. The transitions and the epoch of each observations are labeled on the figure side (top and left respectively). On the right axis of the plot, fluxes are in erg/cm$^2$/s/\AA.
         \label{fig:cenABSev}}
\end{figure*}

For the absorption structures, we see only what is along the line of sight, i.e. any structure that is lying between us and the hot receding surface of the  WD. During maximum and early decline when we see the P-Cyg profiles, the absorbing gas and the pseudo-photosphere illuminating it are contiguous since the latter is part of the ejecta themselves (or their bottom layer) and subtends a large solid angle that intercepts numerous structures along the line of sight. As the pseudo-photosphere recedes, the number of intercepted structures (i.e. clumps) decreases. Our ability to detect them depends on the inclination and the ejecta characteristics (given sufficient S/N and spectral resolution of the data). Ejecta viewed along the symmetry axis will show a larger number of absorption features. The larger the filling factor, the more absorbing structures are intercepted along the line of sight. The same holds for geometrically thicker ejecta.  

V1369 Cen, which is the nova with the best S/N and sampling, shows complex absorptions for the four UV resonance  transitions CII$\lambda$1334\AA, SiIV$\lambda$1393\AA, CIV$\lambda$1548\AA\, and NV$\lambda$1238\AA, whose ions have different ionization potential energy: $\sim$11 eV for C$^{+}$, 33 eV for Si$^{3+}$, 47 eV for C$^{3+}$ and 77 eV for N$^{4+}$. Figure \ref{fig:cenABSev} shows the different  absorption patterns produced by each resonant transition at each epoch. Each ion recombines and absorbs where conditions are suitable for it to do so. At early epochs (day 252 in the figure) the densities are sufficiently high that even low ions such as C$^{2+}$ can recombine to C$^+$ (and therefore absorb from the ground states) up to relatively high velocities, i.e. up to the outskirts of the ejecta. SiIV, whose absorptions are superposed on a blending emission in the figure, are distributed up to -1750 km/s like CII. In contrast, NV and CIV at that same epoch show absorption troughs (very much like H$^0$ or Fe$^{+}$ during the early decline/iron curtain phase): C and N are four and five times ionized everywhere, but only in the innermost portion of the ejecta is the density high enough for them to recombine to C$^{3+}$ and N$^{4+}$. The number of clumps where this can occur are numerous and blending, and only later on (i.e. from day 458 on) do they appear separately. The range of velocity they cover seems to increase, stretching to both lower and higher velocity limits. The appearance of the low velocity structures at later epochs is because they are seen against the stellar continuum while earlier they lie on the emission line and conflate with the emission structures. The structures at the higher velocities (i.e. $\sim -$2000 km/s) appear as soon as the C$^{4+}$ and N$^{5+}$ have recombined to C$^{3+}$ and N$^{4+}$, respectively (the recombination time for N$^{5+}\rightarrow$N$^{4+}$ is slow) and on day 252 the nitrogen is still 5 times ionized at v$_{rad}<-1700$ km/s. The same holds for C and its recombination C$^{4+}\rightarrow$C$^{3+}$, see also the end of Section~\ref{sec:allabs}.
  
This is consistent with the lack of C$^+$ absorptions in the last epoch spectrum: the majority of C is trapped in high ionization states and the fraction of C in low ionization states is not enough to produce a sufficient contrast against the continuum for detection. 
The particular absorption pattern produced by a given ion also depends on the clumps alignment with respect to the changing (decreasing) solid angle subtended by the WD photosphere. Note that most of the structures we see are, in fact, blends even at day 837. The velocity width of individual structures might be as low as 10 km/s, corresponding to radial thickness of only a few percent or less of $R$ (the cone/ejecta maximum radius).

\subsection{Monte Carlo simulation of the absorption features}
\begin{figure}
   \includegraphics[width=9cm]{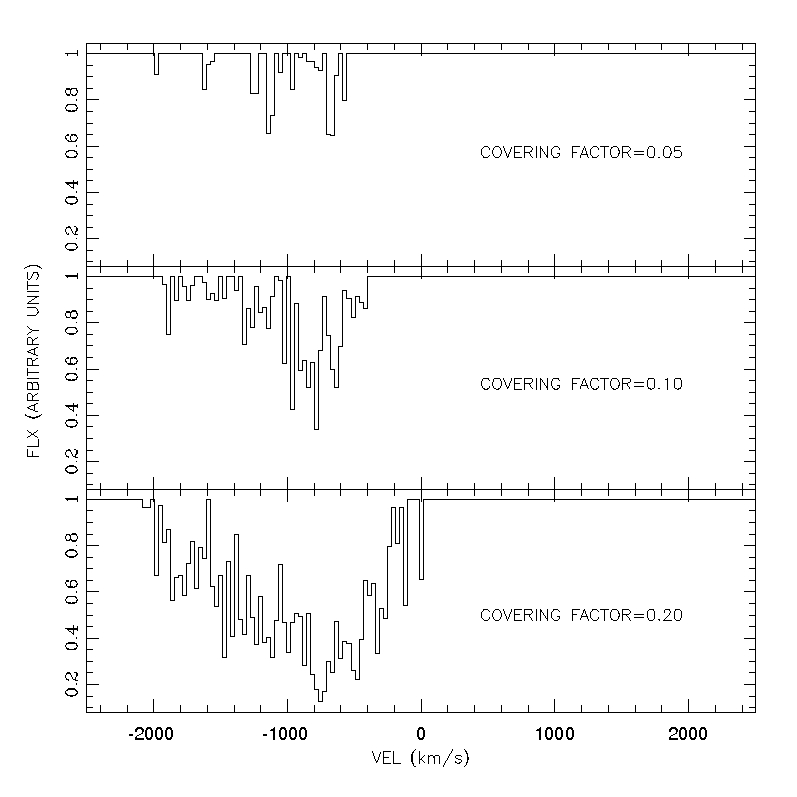}
   \caption{Simulation showing the increasing number of absorption structures for increasing covering factors (see text for more details). 
         \label{fig:mod_a}}
\end{figure}

As a novel feature of our approach we can also check whether the same approximation used to interpret the emission line profile reproduces the absorption features observed in these late stages. Hence, we used the same Monte Carlo code, with the same set of parameters and geometry as for the emission profiles (see Fig.\ref{fig:mod_e} and Section~\ref{sec:sec_mc}), to simulate the absorption features. In this case, each density-point along a selected line of sight is used to compute the opacity. Screening effects between clumps have been ignored, although superposition is included. This approximation is valid for the late nebular phase spectra, since the velocity separation between clumps along the line of sight is larger than the velocity width of the individual clumps. 
The result is shown in Fig.\ref{fig:mod_a}. The figure shows that decreasing the covering factor reduces the number of absorption features, as discussed above.

\subsection{Absorption features in other novae}\label{sec:allabs}

\begin{figure*}
\plotone{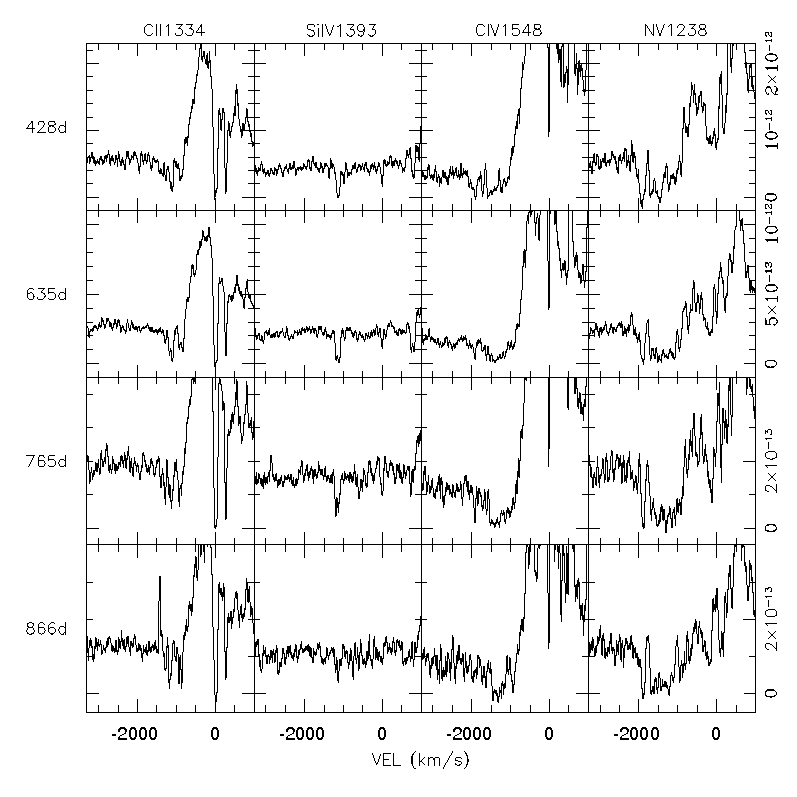}
\caption{V339 Del ionization structures as evinced from the absorptions observed in the permitted resonant transitions. The transitions and the epoch of each observation are labeled on the figure side (top and left respectively). On the right axis of the plot, fluxes are in erg/cm$^2$/s/\AA.
\label{fig:delABSev}}
\end{figure*}

The UV resonance line absorptions were first observed in T~Pyx (day 834 spectrum, Shore et al. 2013a) in the very late observations of the recurrent nova and only in  NV$\lambda$1238\AA\, and CIV$\lambda$1548\AA\, absorptions but with only two structures per transitions. This is consistent with a lower mass for the T~Pyx ejecta and with a low filling factor. Unfortunately, we do not have late UV spectra of V959 Mon, so we cannot say anything about its absorption structures in the UV resonance transition. The V339 Del absorptions are shown in Fig.\ref{fig:delABSev}. The overall evolution of the absorptions in the four resonance transitions is very similar to V1369 Cen, but with a few notable differences. 1) The CII$\lambda$1334\AA\, absorptions persist to the end, unlike V1369 Cen, indicating that at similar epochs V1369 Cen reaches lower densities than V339 Del, too low for C$^{2+}$ to recombine to C$^+$ (the recombination cross section of C$^{2+}$ is smaller than that of C$^{3+}$ hence is favored in higher density environments). 2) The CIV and NV absorptions in V339 Del are never as fragmented as in V1369 Cen, indicating a different distribution of the clumps along the line of sight (the cumulated EW are comparable in the two novae).  

\begin{figure}
\gridline{\fig{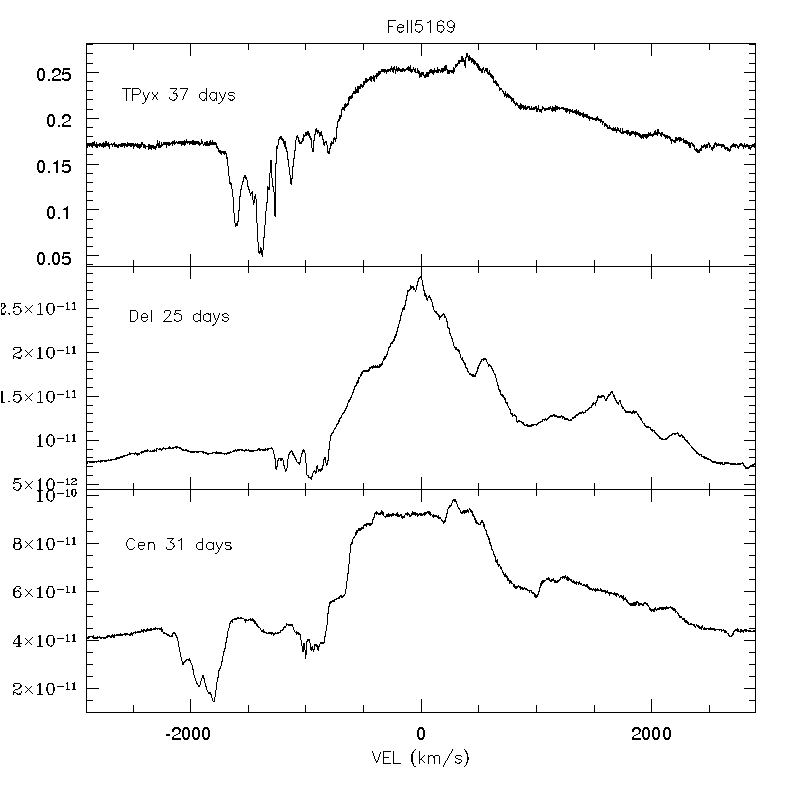}{0.5\textwidth}{}}
\gridline{\fig{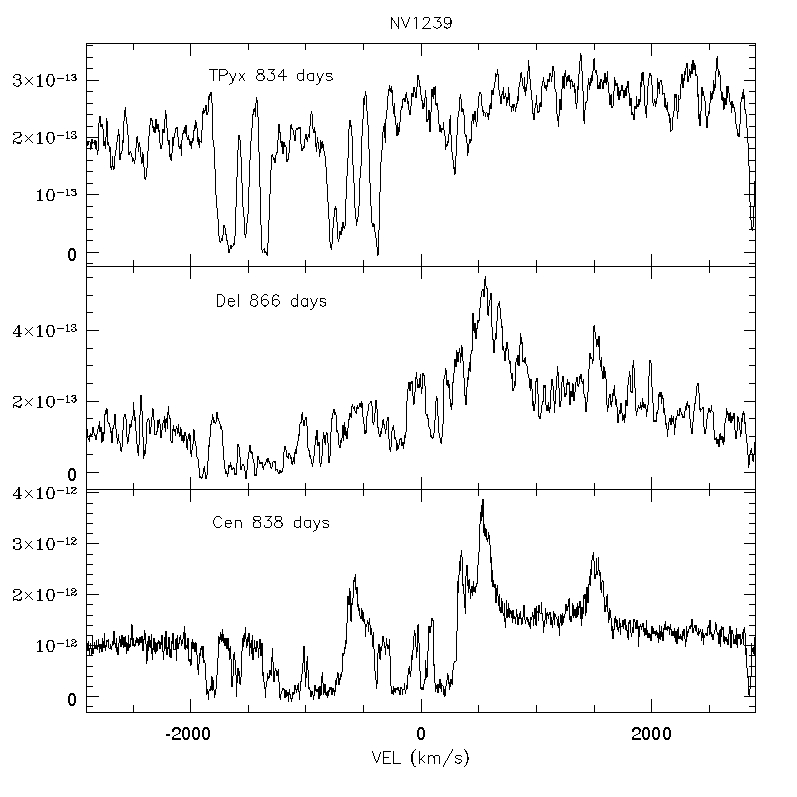}{0.5\textwidth}{}}
\caption{Absorption structures in the spectra of our ``panchromatic novae'' during early decline in the FeII$\lambda$5169\AA\, line (top), and during late nebular phase in the NV$\lambda$1239 \AA\, transition (bottom). 
         \label{fig:abs_comp}}
\end{figure}

Global commonalities that we should note are: i) the velocity ranges spanned by the absorptions are  comparable among novae, and ii) a number of absorption structures persist at the same velocity through all epochs. To better illustrate this, we plot in Fig.\ref{fig:abs_comp} the absorption structures observed in the low ionization potential metals at early time and in the high ionization energy ions at late epochs for all the novae studied in panchromatic high resolution spectroscopy. The figure shows that structures are common to all of them at early decline as soon as the pseudo-photosphere has begun to recede and that similar structures are still visible at very late epochs once the ejecta is frozen in a high ionization state. There is no one-to-one correspondence of the structures at the two epochs for 1) the opacity of the ejecta is very different and at early phases is rapidly varying as the ejecta expand; 2) the photosphere against which the absorptions are detected is different at the two epochs and is, in particular, much smaller (being no longer within the ejecta but matching instead the bloated WD photosphere) in late nebular phase. Nonetheless, one can notice that the structure visible at -1450 km/s in the T Pyx 37 day spectrum is the same as the 834 day spectrum. More generally, where we detect bands of absorbing structures at early epochs we detect absorbing structures later on. The largest velocities observed at the two epochs are comparable to those observed in the wings of the emission lines when their emission measure was sufficient. We are, therefore, always seeing the same unchanged distribution of clumps under different illumination, i.e.  ionization degrees compatible with their local density and the light source underneath. Their absolute position in space is stretching linearly with time as the ejecta expand. This matches a Hubble flow, i.e. the structures are in ballistic expansion.
 
Another important feature of ballistic expansion is that the timescale is unique, independent of the radial distance within the ejecta.  Since the recombination time depends on the electron density as $n_e^{-1}$, and the electron density varies as $n_e \sim r^{-3}\sim$v$_{rad}^{-3}$, at any epoch after the ejection there is a unique radial velocity (in the observer's frame) that corresponds to that density.  It is an expected consequence that the dominant ionization stage of any species is confined at any time in a range of velocities that depends only on the maximum velocity of the ejecta and its mass.  Hence, the radial velocity interval in which the discrete absorption features will be observed for any ion will always be about the same, among novae, even if the individual structures are different.  For example, for $n_e \sim 10^5$cm$^{-3}$, as we have in the last epoch of V1369 Cen, the NV and CIV should be observed at around v$_{rad} \approx$ -1500 km s$^{-1}$.  

\section{Discussion}\label{sec:discussion}
\subsection{Commonalities among novae: the model from the UV+optical high resolution spectroscopic monitoring}

\begin{figure*}
   \plotone{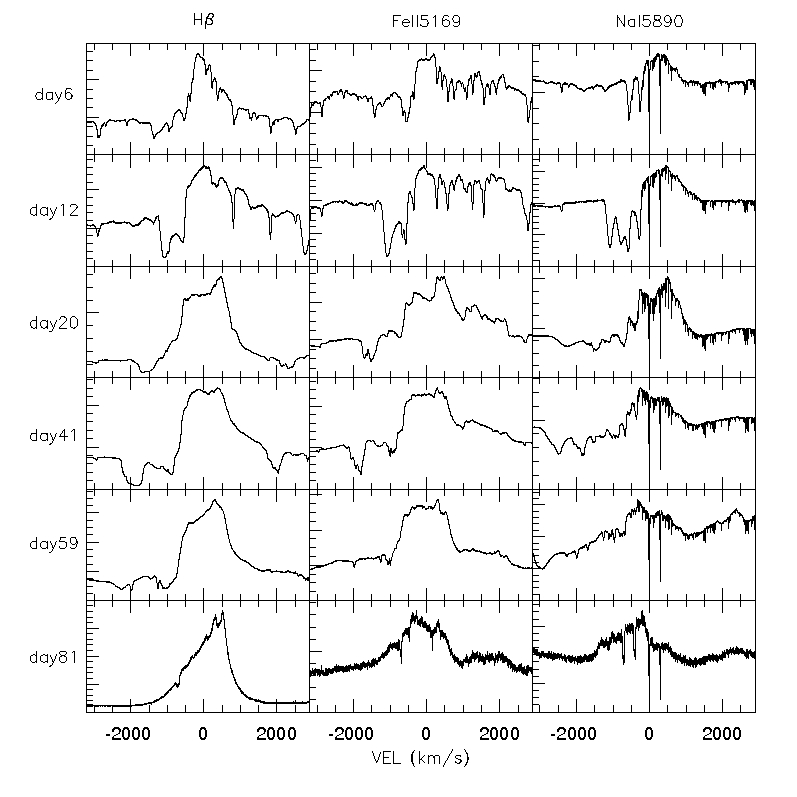}
   \caption{V1369 Cen early decline line profiles. 
         \label{fig:cen_early}}
\end{figure*}

The observational fact emerging from the analysis of the data-sets presented in this and the previous works of the series is the presence of persistent absorption structures which do not move, in velocity space, with time.  This is evident when comparing late nebular spectra of a same nova (e.g. Figure 30 in De Gennaro Aquino et al. 2014; and Fig.\ref{fig:cenABSev} and Fig.\ref{fig:delABSev} in this paper). The persistence of the absorption structures is also shown by comparison of the spectra taken during early decline with those taken during late nebular stages (e.g. Fig.\ref{fig:abs_comp}), and by the evolution of the line profiles across time (e.g. Fig.\ref{fig:cen_early} and Figures 2 and 4 in Shore et al. 2011; Figure 2 and 3 in Shore et al. 2016). The absorptions from low ionization energy ions visible in a given velocity interval at early stages are also present at later epochs in the resonance transitions of high ionization energy ions. They indicate  clumps that have only changed their ionization state, while following the global expansion of the ejecta. The absorption troughs observed during the initial phases of the outburst break into complex absorption structures and tend to shift toward higher v$_{rad}$. This matches a clumpy ejecta whose outer portions cool and recombine as the ejecta expand, and as the pseudo-photosphere recedes, the line of sight intercepts progressively fewer structures. Note that these absorptions can also display apparent inward motion or oscillations (e.g. McLaughlin 1957; Figure 1 in Williams et al. 2008; our Fig.\ref{fig:cen_early}) depending on the nova characteristics  and the cadence of the monitoring: each clump absorbs (at a given wavelength) when its density and the incident radiation from behind allow it. 

Persistent structures that do not move in velocity space are, by definition, a Hubble/radial flow (or in ballistic expansion). {\it That they are stationary at velocities well below the maximum velocity observed during early phases of the expansion is not compatible with a wind} (e.g. Lamers \& Cassinelli 1999). Therefore, a major conclusion of our works is that, in the formation of the observable ejecta, there is no wind ejection at any phase and the ejecta are expanding undisturbed. 

In the past, the absorption features observed at early stages in view of their apparent acceleration (despite some exceptions) were interpreted as subsequent shells ejected with increasing velocity. Were that the case we would not be observing the same structures years after the outburst. 
It is interesting to note that Liimets et al. (2012) reach similar conclusions (ballistic expansion and little or no deceleration) analyzing the resolved clumps over the last 25 yr of the $>$100 yr old nova GK Per. 

Our monitoring shows that the measured line flux and the derived density in the late nebular spectra of our data sets match the evolution expected for a ballistic expansion. This same conclusion was previously reached by Shore et al. (1996) by analyzing the UV line flux of V1974 Cygni together with its x-ray light curve. 

An undisturbed ballistic expansion is incompatible with multiple ejections of shells or structures having increasing velocity (e.g. McLaughlin 1957, Friedjung 1987, Kimeswenger et al. 2008, Tanaka et al. 2011, Chomiuk et al 2014, Arai et al. 2016). In addition, the velocity field of a ballistic expansion is incompatible with any geometry that implies acceleration or deceleration of the ejecta (e.g. an hourglass geometry). 

It is known from imaging that nova ejecta are not spherical but elongated (e.g. Slavin et al. 1995; Gill \& O'Brien 2000; Harman \& O'Brien 2003; Liimets et al. 2012; Shara et al. 2015; Hjellming 1995; Chomiuk et al. 2014; Lindford et al. 2015; Healy et al. 2017; Lane et al. 2007; Chesneau et al. 2011; Schaefer et al. 2014), yet the exact geometry is far from certain. The simplest geometry compatible with a non spherical radial expansion is polar caps or a biconical form (the difference between the two depending on the thickness one considers). 
Combining this consideration with the clumpiness of the ejecta, we used a simple Monte Carlo code to mimic the observed line profiles. We adopt a biconical geometry with random clumps that obey a velocity and density distribution matching the ballistic expansion, and emissivity from recombination as appropriate for the late nebular spectra.  
Our series of papers has shown that such a geometry suffices in reproducing the gross line profiles we observe in high resolution spectroscopy. Therefore, at this stage, there is no need to consider more complex geometries. The detailed reproduction of individual features would require real dynamical models which are beyond the scope of this series. 

An interesting test of the validity of our approach comes from the use of the adopted geometry in the computation of the ejecta mass and filling factor. They are computed adopting the identified emitting volume and integrating the $n\sim$1/r$^3$ density within such volume. Each epoch is treated independently providing ejecta mass estimates that are the same within the uncertainties (e.g. Figure 14 of Shore et al. 2016). 

The observed spectral evolution is a further validation of the proposed picture. 
At late phases the highest ions sample the outermost regions of the ejecta because the lower density there prevents recombination. Neutral or singly ionized transitions are favored in the densest region of the ejecta and are expected to be observable for several years in the inner regions of the most massive ejecta. Were the density following a different distribution we would not have observed such an ionization structure.   

\subsection{The scenario at other wavelengths: a qualitative test}

Observations by the Fermi/LAT satellite have revealed that novae are capable of emitting $\gamma$-rays with energy $\geq$100 MeV that are evidence of shocks among accelerated particles ($Fermi$ collaboration 2014; Cheung et al. 2016). 
Metzger et al. (2014) propose that the 100 MeV emission arises from the collision of distinct ejections of gas having mass differences within an order of magnitude and relative velocity in the range 0.15-0.60. These conditions match the observed range of velocities (from a few hundreds to a few thousands km/s; Figs~\ref{fig:cenABSev}, \ref{fig:delABSev} and \ref{fig:abs_comp}; see also Figures 29 and 30 of De Gennaro Aquino et al. 2014) and the measured local projected densities at any given time (e.g. Fig.~\ref{fig:cen_ne}). Hence, collisions between clumps during the ejection and the initial expansion could explain the 100 MeV emission. More than that, intra-clump collisions are expected in a single short duration explosion that generates a stochastic range of clump velocities. The orientation and porosity of the ejecta and the density of individual clumps would then power the $\gamma$ and x-ray light curves. The modeling of those curves, like precise line profile fitting, would require assuming clump distributions which are beyond the scope of this paper. 

Centimeter wavelength radio observations have shown morphological changes of the ejecta in resolved images (e.g. Hjellming 1996 and references therein; Chomiuk et al. 2014; Wendeln et al. 2017). Different morphologies observed in different epochs at different frequencies can be explained by differential evolution of the opacity in the spatially resolved non spherical inhomogeneous sources, similarly to the changes in the line profiles during early stages of the outburst. Indeed, the ``varying'' radio images during the optically thick phase of the novae were explained by non-spherical single ejecta having a velocity gradient (e.g. Hjellming 1996 and 1995).  Global density gradients and inhomogeneous ejecta were instead inferred from radio observations deviating from to the theoretical distribution expected for a uniform optically thick and isothermal gas (e.g. Seaquist \& Palimaka 1977). Morphological difference in resolved radio images of an optically thin gas (e.g. Chomiuck et al. 2014), maps the different emissivity of each region at a given frequency, similarly to the profile difference we observe in the optical in the late nebular phase (e.g. Fig.~\ref{fig:masters} bottom panel, compare NIV] to H$\beta$ for an example of a dramatic difference). 

\section{Summary and conclusion}
In this paper we have analyzed V1369 Cen nebular spectra and compared them with those of the novae observed with similar instrumentation in order to both determine the physical parameters of V1369 Cen ejecta and spot commonalities among novae that could allow the identification of a unified picture for their ejecta and ejecta kinematics. 

Our quantitative results for the physical parameters of V1369 Cen are: 
1) E(B-V)=0.15 mag; 2) distance in the range 1.8--2.4 kpc; 3) filling factor in the range 0.1--0.2; 4) ejecta mass $\sim$1$\times$10$^{-4}$ M$\odot$; 5) anomalous Ne abundance and anomalously large N/C abundance ratio. 

Our effort toward a unified picture for nova ejecta is necessarily qualitative. It makes use of a single assumption (the single explosion event), and of the evidence that the ejecta are neither spherical nor homogeneous. 
The presence of stationary structures is, instead, new observational evidence from our data. Their interpretation requires that they constitute a ballistic expansion, contradicting any picture of wind-like ejecta. 
The observed kinematics automatically define a density distribution and, therefore, behaviors that can be verified (and/or predicted) through the monitoring. Examples are: the outward propagation of a recombination wave during the early phase of the ejecta expansion (the iron curtain phase), the higher ionization degree being larger in the outskirts of the ejecta, the decline of the emission line intensity once the SSS has turned off and the ejecta are dominated by recombination, the frozen line profile and ionization structure, again, once the ejecta cool mainly by recombination. 
The biconical geometry globally constraints the ensemble of structures in the aspherical ejecta to reproduce the gross observable features (line width, maximum velocity, more or less ``rectangular'' or ``saddle shaped'' emission line profiles). 
The resulting model, beside being able to explain the panchromatic high resolution spectroscopic monitoring, is consistent with recent observations at other wavebands (e.g. $\gamma$-ray and radio). On the contrary, other proposed scenarios (e.g. colliding winds and shells), while capable of explaining those wavebands and/or individual observations or short term monitoring, are incompatible with the data sets shown for V959 Mon, V339 Del, V1369 Cen and even T~Pyx in our series of papers.

We conclude noting that, while we have achieved a self-consistent picture for the observable nova ejecta, the formation of the ejecta structures remains an open problem. 

\acknowledgments

Based on observations made with:  1) the NASA/ESA Hubble Space Telescope, obtained from the data archive at the Space Telescope Science Institute. STScI is operated by the Association of Universities for Research in Astronomy, Inc. under NASA contract NAS 5-26555; 
 2) the Nordic Optical Telescope, operated on the island of La Palma jointly by Denmark, Finland, Iceland, Norway, and Sweden, in the Spanish Observatorio del Roque de los Muchachos of the Instituto de Astrofisica de Canarias; 3) the ESO Very Large Telescope at the Paranal observatory; and 4) the ESO/MPG 2.2m in the La Silla observatory.  

The authors are grateful to Teddy Cheung, Jordi Jos\'e, Jan-Uwe Ness, Julian Osborne, Francois Teyssier and Brian Warner for having participated to the 2013 Pisa meeting, helping defining this research program. 
EM warmly thanks Pierluigi Selvelli for the enlightening spectroscopic confrontations and the friendly ``calories support''. SNS thanks Thomas Augusteijn, Bob Gherz, Pierre Jean, Paul Kuin, Sumner Starrfield, Fred Walter and Laura Chomiuk for discussions and John Telting for his help with the NOT observations and the Astronomical Institute of the Charles University for a visiting professorship.
The authors thank the anonymous referee for the helpful critiques. 

%

\vspace{5mm}
\facilities{HST(STIS), VLT(UVES), NOT(FIES), Swift(XRT), AAVSO}


\software{IRAF, Super Mongo, IDL  
          }



\appendix

\section{Identified lines and their profile characteristics}
The use of high resolution spectroscopy has not only revolutionized our understanding of the ejecta geometry, structure and kinematics but also our ability to identify atomic species and their ions. It is possible to identify the species forming a blend by comparing peaks or structures and wings of the line profile. When there still is ambiguity on the transition identification, the profile characteristics, if discernible, can be used to solve the ambiguity since ions of similarly ionization potential energy produce  transitions of similar profile. In addition, the ability of distinguishing different profile types allows us to recognize when a transition is ``replaced'' by another due to the change of ionization state of the ejecta. 
In the figure below we present a sample of ``master'' lines for the first (day 252, top panel) and the last (day 836, bottom panel) epoch.  Those profiles were used to help the line identification listed in Table~\ref{tab:obs_lines}, the observed lines are listed together with their profile types. Note that the profile types at the two epochs are independent because the emission measure of a given transition changes with time affecting the profile. Hence, each set of profile types refers exclusively to the epoch in which it is used. 

\startlongtable
\begin{deluxetable}{cccc}\label{tab:obs_lines}
\tablecaption{Observed transitions and their line profile typology (cross check with profiles of the corresponding epoch in Fig.\ref{fig:masters}) for the UV range of the spectrum. The ``:'' notation is applied to those measurements or identification or profile assignment which are uncertain. The ``b'' assignment means that there is in blend, while the ``w'' indicated weak lines.  }
\scriptsize
\tablehead{
\colhead{{\bf $\lambda$ (\AA)}}& \colhead{{\bf transition}} & \colhead{{\bf day 252 profile}} & \colhead{{\bf day 838 profile}}
}
\startdata
1239, 1243 & NV & B & D \\
1302, 1305 & OI & w & - \\
1323 & ? & A & -\\
1336, 1334 & CII & A: & D\\
1343 & OIII & A & - \\
1355 & OI: & A & - \\
1371 & OV & B & A: \\
1389 & ? & A & - \\
1401, 1405 & OIV & B & D \\
1411 & ? & A/B & - \\
1486 & NIV & B & D \\
1548, 1551 & CIV & B & D\\
1526, 1529 & ? & A: & - \\
1575 & [NeV] & C: & - \\
1601 & [NeIV] & B & - \\
1619 & CIII & B & - \\
1640 & HeII & B & B \\
1661, 1666 & OIII & B & w \\
1718 & NIV & B & - \\
1749, 1751 & NIII & B: & - \\
1884, 1892 & SiIII] & w & - \\
1907, 1909 & CIII & B & D \\
2139, 2143 & NII & A: & - \\
2188 & ? & A: & - \\
2297 & CIII & B & w \\
2327, 2328 & CII & A: & - \\
2344 & ?  & B: & - \\
2470 & OII & A & wD \\
2647 & NIV & B &  - \\
2735 & ?  & B & - \\
2784 & ?  &  B &  wD: \\
2796, 2803 & MgII & A: &  -\\
2827 & ?  & ?  & - \\
2837 & CII: & B & w \\
3049 & ? & B & - \\
\enddata
\end{deluxetable}

\newpage

\begin{figure}
   \gridline{\fig{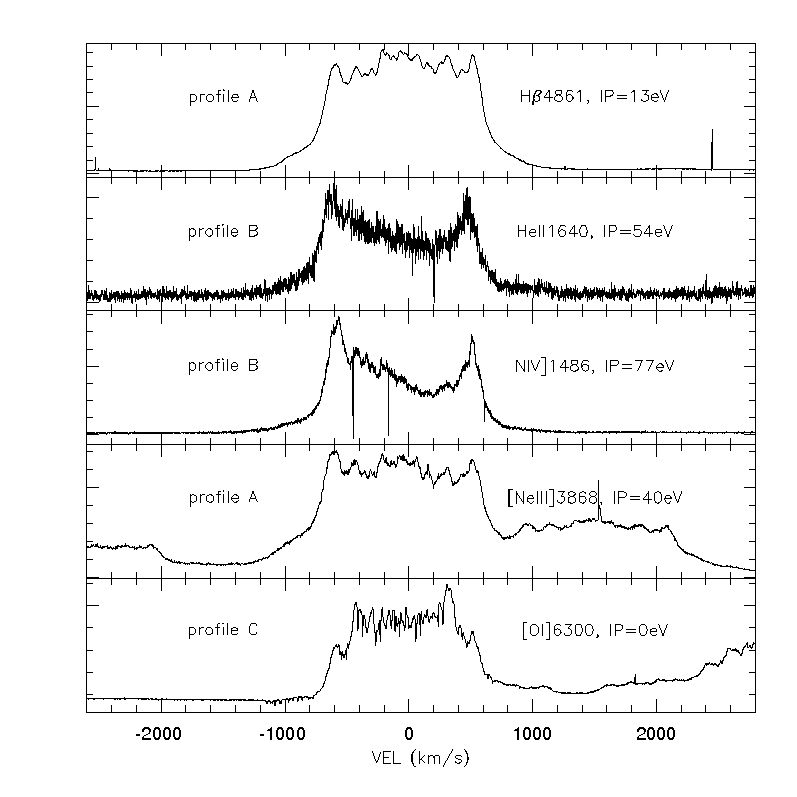}{0.5\textwidth}{}}
   \gridline{\fig{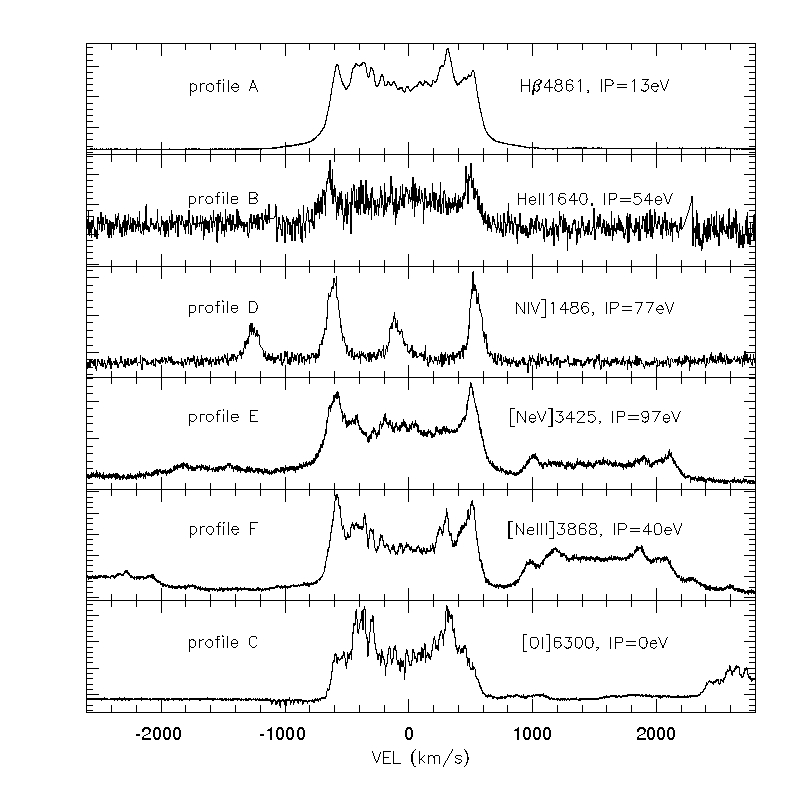}{0.5\textwidth}{}}
   \caption{Master reference profiles used to aid the line identification in two illustrative epochs (top: day 252, bottom: day 836) of V1369 Cen spectral evolution.  
         \label{fig:masters} }
\end{figure}

\newpage

\startlongtable
\begin{deluxetable}{cccc}\label{tab:obs_lines_vis}
\tablecaption{As Table~\ref{tab:obs_lines} but for the visible band.}
\scriptsize
\tablehead{
\colhead{{\bf $\lambda$ (\AA)}}& \colhead{{\bf transition}} & \colhead{{\bf day 252 profile}} & \colhead{{\bf day 838 profile}} 
}
\startdata
3760 & OIII & B & bB/E \\
3771 & H$_{11}$ & b & bA \\
3798 & H$_{10}$ & A & bA \\
3835 & H$\eta$ & A & A \\
3869 & [NeIII] & A & F \\
3889 & H$\zeta$ & A & A \\
3893, 3896 & [FeV] & - & b \\
3968+3970 & [NeIII]+H$\varepsilon$ & b & b \\
3995+4004: & NII+NIII & A & b  \\
4026 & HeI & bA & [FeV]4027, b  \\
4041 & NII & A & b \\
4072, 4076 & OII & b & [FeV]4072, b \\
4102+4097, 4104 & H$\gamma$+NIII & b & b \\
4144+4152 & HeI+CIII & w & w \\
4181:+4190+4196, 4200 & ?+OII+NIII & b & b \\
4237, 4242 & NII & b & b \\
4267, 4274:+4287: & CII+? & bA: & bA \\
4320 & OII & b &  w \\
4341 & H$\gamma$ & A & A \\
4363 & [OIII] & A & F \\
4379 & NIII & A: & B \\
4415, 4417 & OII:  & b & w \\
4447+4448 & NII+OII & b &  \\
4472 & HeI & A & A \\
4511/15/24/31/35+4542 & NIII+HeII & b & b \\
4591 & OII & b: & b \\
4603+4606 & NV:+? & b & 4607?, bA: \\
4634.2, 4640.6+4649 & NIII+OII & b & bB \\
4658 & [FeIII]: & b & b \\
4686 & HeII & A/B & B\\
4702:+4706: & OII+? & b & +4699, b \\
4714,4725 & [NeIV] & b & - \\
4740 & [ArIV] & - & B \\ 
4803 & CII & w & A/C \\
4861 & H$\beta$ & A & A \\
4894 &  [FeVII] & - & wB: \\
4959 & [OIII] & A & F \\
5007 & [OIII] & A & F \\
5041, 5056 & SiII: & b & bB: \\
5147+5158 & [FeVI]+[FeVII] & b & [FeVI], bB: \\
5176 & [FeVI]  & - & wB \\
5309 & [CaV] & B & E \\
5336 & [FeVI] & - & E: \\
5412 & HeII & A & +[FeVI]5425, b \\
5486 & [FeVI] & - & wB \\
5530+5534 & NII+[ArX] & b & w \\
5594 5615 & ? & B  & OIII5592, A/B \\
5632 & [FeVI] &  & wB/E \\
5668, 5676, 5679.6 & NII & b & b+[FeVI]: \\
5721 & [FeVII] & B & B/E \\
5755 & [NII] & C & - \\
5801, 5812 & CIV & b & - \\
5876 & HeI & A & A \\
5932, 5942 & NII & A & w \\
6086 & [CaV] & B &  E\\
6300 & [OI] & C &  C\\
6364+6346, 6371 & [OI]+SiII & b &  [OI], C\\
6482+6460:+6501 & NII+? & bA & w \\
6563 & H$\alpha$+[NII] & bA & [NII], A/C \\
6675 & HeI & A & A \\
6747 & ? & A/B: & w \\
7006 & [ArV] & A/B & B/E \\
7063 & HeI & A & A \\
7136+7150/2 & [ArIII]+OII: & b & [Ar] F\\
7232+7236 & CII, [ArIV] too? & A & [ArIV], A \\
7320, 7331 & [OII] & b & b \\
7705+7717+7726 & NII]: or NIII]: & b & [ArIII]:, 7704+7714?+7751? \\
7774 & OI & b & - \\
8237+8196: & HeII+CIII? & b & HeII,  A: \\
8446 & OI & C & w \\
8545 & Pa$_{14}$ & A & $\dagger$ \\
8598 & Pa$_{13}$ & A & $\dagger$ \\
8665+8680/3: & Pa$_{14}$+NI: & b & $\star$ \\
8750 & Pa$_{12}$ & A & bA \\
8861 & Pa$_{11}$ & $\ddagger$ & A \\
8926 & ? & A & - \\
9015 & Pa$_{10}$ & A & A \\
9068/70 & [SIII]: & bA & [SIII], A\\
\enddata
\tablecomments{\\
$\dagger$ in the gap of the UVES red CCD mosaic.\\
$\star$ partially in the UVES red CCD mosaic.\\
$\ddagger$ partially in the gap between the FEROS orders. }
\end{deluxetable}

\newpage

\startlongtable
\begin{deluxetable}{ccc}\label{tab:obs_lines_vis2}
\tablecaption{Continuation of Table~\ref{tab:obs_lines_vis} for the wavelength range covered by UVES (day 838 observation).   }
\scriptsize
\tablehead{
\colhead{{\bf $\lambda$ (\AA)}} & \colhead{{\bf transition}} & \colhead{{\bf day 838 profile}} 
}
\startdata
3133 & OIII & F/B \\
32033 & HeII & B \\
3266, 32616 & OIII &  bB \\
3312, 3299 & OIII & bB \\
3341+3346 & OIII+[NeV] & bB \\
3381, 3386 & OIV & bB \\
3404, 3412+3426  & OIV+[NeV] & bE \\
3444 & OIII & B \\
3479: & NIV & w \\
3587: & [FeVII] & w \\
3703 & OIII & wb \\
3726+3729 & OIV+[OII] & b \\
3750+3755 & H$_{12}$+OIII  & b \\
 &  &   \\
9229.0 & Pa$_{9}$ & A \\
9532.1 & [SIII] & A \\
9903.5 & ? & w \\
10049.6 & Pa$_{8}$ & A \\
10123.6 & HeII & B \\
10395.6, 10404.1 & [NI] & w \\
\enddata
\end{deluxetable}


\begin{thebibliography}{}
\bibitem[Corrales(2015)]{2015ApJ...805..23C} Arai, A.; Kawakita, H.; Shinnaka, Y.; et al. 2016, AJ, 830, 30
\bibitem[Corrales(2015)]{2015ApJ...805..23C} Asplund, M.; Grevesse, N.; Sauval, A. J.; et al,; 2009, ARA\&A, 47, 481
\bibitem[Corrales(2015)]{2015ApJ...805..23C} Banerjee, D. P. K.; Srivastava, Mudit K.; Ashok, N. M.; et al.; 2016, MNRAS, 455, 109
\bibitem[Corrales(2015)]{2015ApJ...805..23C} Cardelli, J.A.; Clayton, G.C.; Mathis, J.S.; 1989, ApJ, 345, 245
\bibitem[Corrales(2015)]{2015ApJ...805..23C} Casanova, J.; Jos\'e, J.; García-Berro, E.; Shore, S. N.; 2016, A\&A, 595, 28
\bibitem[Corrales(2015)]{2015ApJ...805..23C} Cassatella, A.; Lamers, H. J. G. L. M.; Rossi, C.; et al., 2004a, A\&A, 420, 571
\bibitem[Corrales(2015)]{2015ApJ...805..23C} Cassatella, A.; Gonzales-Riestra, R.; Selvelli, P.; 2004b, ``Classical novae'', INES access guide N.3, ESA Publication Division
\bibitem[Corrales(2015)]{2015ApJ...805..23C} Cheung, C. C.; Jean, P.; Shore, S. N.; et al.; 2016, ApJ, 826, 142 
\bibitem[Corrales(2015)]{2015ApJ...805..23C} Chesneau, O.; Meilland, A.; Banerjee, D. P. K.; et al.; 2011. A\&A, 534, 11
\bibitem[Corrales(2015)]{2015ApJ...805..23C} Chochol, D.; Grygar, J.; Pribulla, T.; et al.; 1997, A\&A, 318, 908
\bibitem[Corrales(2015)]{2015ApJ...805..23C} Chomiuk, L.; Linford, J. D.; Yang, J.; et al.; 2014, Nat, 514, 339
\bibitem[Corrales(2015)]{2015ApJ...805..23C} De Gennaro Aquino, I.; Shore, S. N.; Schwarz, G. J.; et al. 2014, A\&A, 562, A28 	
\bibitem[Corrales(2015)]{2015ApJ...805..23C} Downen, L. N.; Iliadis, C.; Jos\'e, J.; Starrfield, S,; 2013, ApJ, 762, 105
\bibitem[Corrales(2015)]{2015ApJ...805..23C} {\it Fermi} Collaboration; 2014, Sci, 345, 554 
\bibitem[Corrales(2015)]{2015ApJ...805..23C} Eyres, S. P. S.; Heywood, I.; O'Brien, T. J.; et al., 2005, MNRAS, 358, 1019
\bibitem[Corrales(2015)]{2015ApJ...805..23C} Friedjung, M.; 1966a, MNRAS, 131, 447
\bibitem[Corrales(2015)]{2015ApJ...805..23C} Friedjung, M.; 1966b, MNRAS, 132, 143
\bibitem[Corrales(2015)]{2015ApJ...805..23C} Friedjung, M.; 1987, A\&A, 180, 155
\bibitem[Corrales(2015)]{2015ApJ...805..23C} Gill, C. D.; O'Brien, T. J.; 1999, MNRAS, 307, 677
\bibitem[Corrales(2015)]{2015ApJ...805..23C} Gill, C. D.; O'Brien, T. J.; 2000, 314, 175
\bibitem[Corrales(2015)]{2015ApJ...805..23C} Harman, D. J.; O'Brien, T. J., 2003, MNRAS, 344, 1219
\bibitem[Corrales(2015)]{2015ApJ...805..23C} Harvey, E.; Redman, M. P.; Boumis, P.; et al.; 2016, A\&A 595, 64
\bibitem[Corrales(2015)]{2015ApJ...805..23C} Healy, F.; O'Brien, T. J.; Beswick, R.; et al.; 2017, MNRAS, 469, 3976
\bibitem[Corrales(2015)]{2015ApJ...805..23C} Hjellming, R. M.; 1990, LNP (Lecture Notes in Physics), in {\it Physics of classical novae}, Proceeding of Colloquium N. 122 of the IAU, 1990, 369, 169 
\bibitem[Corrales(2015)]{2015ApJ...805..23C} Hjellming, R. M.; 1995, in {\it Cataclysmic Variables}, Proceedings of the conference held in Abano Terme in June 1994, Astrophysics and Space Science Library, 205, 139 
\bibitem[Corrales(2015)]{2015ApJ...805..23C} Hjellming, R. M.; 1996, ASPC, 93, 147
\bibitem[Corrales(2015)]{2015ApJ...805..23C} Hutchings, J. B., 1970, PDAO (Publications of the Dominion Astrophysical Observatory), vol.13, p.397
\bibitem[Corrales(2015)]{2015ApJ...805..23C} Jose', J.; Shore, S. N.; 2008, in {\it Classical novae}, 2$^{nd}$ edition, Cambridge University Press
\bibitem[Corrales(2015)]{2015ApJ...805..23C} Kafka, S.; 2017, Observations from the AAVSO International Database, https://www.aavso.org
\bibitem[Corrales(2015)]{2015ApJ...805..23C} Kalberla, P.M.W.; Burton, W.B.; Hartmann, D.; Arnal, E.M.; et al.; 2005, A\&A, 440, 775
\bibitem[Corrales(2015)]{2015ApJ...805..23C} Kalberla, P.M.W.; Haud, U.; 2015, A\&A, 578, A78 
\bibitem[Corrales(2015)]{2015ApJ...805..23C} Kimeswenger, S.; Dalnodar, S.; Knapp, A.; et al.; 2008, A\&A, 492, 787
\bibitem[Corrales(2015)]{2015ApJ...805..23C} Lamers, H. J. G. L. M.; Cassinelli, J. P.; 1999, {\it Introduction to Stellar Winds}, Cambridge University Press
\bibitem[Corrales(2015)]{2015ApJ...805..23C} Lane, B. F.; Retter, A.; Eisner, J. A.;  et al.; 2007, ApJ, 699, 1150
\bibitem[Corrales(2015)]{2015ApJ...805..23C} Livio, M.; Truran, J. W.; 1994, ApJ, 425, 797 
\bibitem[Corrales(2015)]{2015ApJ...805..23C} Liimets, T.; Corradi, R. L. M.; Santander-García, M.; et al. ApJ, 761, 34 
\bibitem[Corrales(2015)]{2015ApJ...805..23C} Linford, J. D.; Ribeiro, V. A. R. M.; Chomiuk, L.; 2015, ApJ, 805, 136
\bibitem[Corrales(2015)]{2015ApJ...805..23C} McLaughlin, D. B.; 1957, Vistas in Astronomy, vol. 2, issue 1, page 1147
\bibitem[Corrales(2015)]{2015ApJ...805..23C} McLaughlin, D. B.; 1964, AnAp, 27, 450
\bibitem[Corrales(2015)]{2015ApJ...805..23C} Metzger, B.D.; Hasco\"et, R.; Vurm, I.; et al.; 2014, MNRAS, 442, 713
\bibitem[Corrales(2015)]{2015ApJ...805..23C} Munari, U.; Zwitter, T.; 1997, A\&A, 318, 269
\bibitem[Corrales(2015)]{2015ApJ...805..23C} Munari, U.; Ribeiro, V. A. R. M.; Bode, M. F.; et al.; 2011, MNRAS, 410, 525
\bibitem[Corrales(2015)]{2015ApJ...805..23C} Nussbaumer, H.; Schild, H.; 1979, A\&A, 75, L17
\bibitem[Corrales(2015)]{2015ApJ...805..23C} Nussbaumer, H.; Schild, H.; 1981, A\&A. 101, 118
\bibitem[Corrales(2015)]{2015ApJ...805..23C} O'Brien, T. J.; Bode, M. F.; 2008, in {Classical Novae}, 2$^{nd}$ edition, Cambridge University Press
\bibitem[Corrales(2015)]{2015ApJ...805..23C} Page, K. L.; Osborne, J. P.; Wagner, R. M.; et al.; 2013, ApJ, 768, L26
\bibitem[Corrales(2015)]{2015ApJ...805..23C} Payne-Gaposchkin, C.; Menzel, D. H.; 1938, Har. Circ., N.428
\bibitem[Corrales(2015)]{2015ApJ...805..23C} Payne-Gaposchkin, C.; 1957, {\it The Galactic Novae}, North-Holland Publishing Company Amsterdam 
\bibitem[Corrales(2015)]{2015ApJ...805..23C} Ribeiro, V. A. R. M.; Darnley, M. J.; Bode, M. F.; et al.; 2011, MNRAS, 412, 1701
\bibitem[Corrales(2015)]{2015ApJ...805..23C} Ribeiro, V. A. R. M.; Munari, U.; Valisa, P 2013ApJ...768...49R
\bibitem[Corrales(2015)]{2015ApJ...805..23C} Schwarz, G. J.; 2014, {\it Stella Novae: Past and Future Decades}, ASPC, 490, 101
\bibitem[Corrales(2015)]{2015ApJ...805..23C} Scott, A.D.; Duerbeck, H.W.; Evans, A.; et al.; 1995, A\&A 296, 439
\bibitem[Corrales(2015)]{2015ApJ...805..23C} Seaquist, E. R.; Palimaka, J.; 1977, ApJ, 217, 781
\bibitem[Corrales(2015)]{2015ApJ...805..23C} Shara, Michael M.; Zurek, D.; Schaefer, B. E.; et al.; 2015, ApJ, 805, 148 
\bibitem[Corrales(2015)]{2015ApJ...805..23C} Schaefer, G. H.; Brummelaar, T. T.; Gies, D. R.; et al.; 2014, Nature, 515, 234
\bibitem[Corrales(2015)]{2015ApJ...805..23C} Shore, S. N.; Starrfield, S.; Sonneborn, G.; 1996, ApJ, 463, L21
\bibitem[Corrales(2015)]{2015ApJ...805..23C} Shore, S. N.; Augusteijn, T.; Ederoclite, A.; et al., 2011, A\&A, 533, L8
\bibitem[Corrales(2015)]{2015ApJ...805..23C} Shore, S. N.; Schwarz, G. J.; De Gennaro Aquino, I.; et al.,  2013a ,A\&A, 549, 140
\bibitem[Corrales(2015)]{2015ApJ...805..23C} Shore, S. N.; De Gennaro Aquino, I.; Schwarz, G. J.; et al., 2013b, A\&A, 553, 123
\bibitem[Corrales(2015)]{2015ApJ...805..23C} Shore, S. N.; 2013c, A\&A, 559, L7
\bibitem[Corrales(2015)]{2015ApJ...805..23C} Shore, S. N.; Mason, E.; Schwarz, G. J.; et al.; 2016, A\&A, 590, 123
\bibitem[Corrales(2015)]{2015ApJ...805..23C} Slavin, A. J.; O'Brien, T. J.; Dunlop, J. S.; 1995, MNRAS, 276, 353
\bibitem[Corrales(2015)]{2015ApJ...805..23C} Starrfield, S.; Iliadis, C.; Hix, W. R.; 2016, PASP, 128, 1001
\bibitem[Corrales(2015)]{2015ApJ...805..23C} Tanaka, J.; Nogami, D., Fujii, M.; et al.; 2011, PASJ, 63, 159
\bibitem[Corrales(2015)]{2015ApJ...805..23C} Vanlandingham, K. M.; Schwarz, G. J.; Shore, S. N.; et al.; 2001, AJ, 121, 1126
\bibitem[Corrales(2015)]{2015ApJ...805..23C} Wendeln, C.; Chomiuk, L.; Finzell, T.; et al.; 2017; ApJ, 840, 110
\bibitem[Corrales(2015)]{2015ApJ...805..23C} Williams, P. M.; Longmore, A. J.; Geballe, T. R.; 1996, MNRAS, 279, 804
\bibitem[Corrales(2015)]{2015ApJ...805..23C} Williams, R. E.; 1992, AJ, 104, 725
\bibitem[Corrales(2015)]{2015ApJ...805..23C} Williams, R. E.; Mason, E.; Della Valle, M.; et al.; 2008, ApJ, 685, 451

\end{thebibliography}
\end{document}